\documentclass[acmsmall,
screen,
authorversion=true,
nonacm=true,]{acmart}

\AtBeginDocument{%
  }

\usepackage[ruled,vlined]{algorithm2e} 
\usepackage{subfigure} 
\usepackage{enumitem} 
\usepackage{pgfplots} 
\pgfplotsset{compat=1.6}
\usepackage{pgfplotstable}
\pgfplotsset{compat=1.6}
\usepackage{xcolor}

\usepackage{graphicx} 

\usepackage{wrapfig}

\usepackage{subcaption}

\usepackage{amsfonts}
\usepackage{mathtools,commath,nicefrac}

\usepackage{tabularx}
\usepackage{booktabs}

\usepackage[frozencache=true,cachedir=minted-cache]{minted}

\usepackage{enumitem}
\setlist[itemize]{leftmargin=*}
\setlist[enumerate]{leftmargin=*}
\usepackage{tikz}
\usepackage{tikzpeople}

\usepackage{pgfplots}
\usepackage{comment}
\usepackage[obeyclassoptions,mode=buildnew]{standalone}

\usepackage{listings}
\usepackage[ruled,vlined]{algorithm2e}

\usepackage{makecell}

\usepackage{diagbox}

\usepackage{fontawesome}

\newcommand{\ceil}[1]{\lceil #1 \rceil}

\newcommand{\decodable }[1][d]{$#1$-decodable}

\newcommand{\LFFZ}[2][n]{$\del{#1,#2}$-LFFZ}

\newcommand{\RNum}[1]{\uppercase\expandafter{\romannumeral #1\relax}}

\newcommand{\rNum}[1]{\lowercase\expandafter{\romannumeral #1\relax}}

\definecolor{myblue}{RGB}{80,80,160}
\definecolor{mygreen}{RGB}{80,160,80}

\usetikzlibrary{shapes,decorations,shadows,positioning,chains,fit,shapes,calc,matrix,backgrounds,arrows}

\pgfplotsset{compat=1.18}
\usetikzlibrary{svg.path}

\makeatletter
\renewcommand{\SetKwInOut}[2]{%
  \sbox\algocf@inoutbox{\KwSty{#2}\algocf@typo:}%
  \expandafter\ifx\csname InOutSizeDefined\endcsname\relax
    \newcommand\InOutSizeDefined{}\setlength{\inoutsize}{\wd\algocf@inoutbox}%
    \sbox\algocf@inoutbox{\parbox[t]{\inoutsize}{\KwSty{#2}\algocf@typo:\hfill}~}\setlength{\inoutindent}{\wd\algocf@inoutbox}%
  \else
    \ifdim\wd\algocf@inoutbox>\inoutsize%
    \setlength{\inoutsize}{\wd\algocf@inoutbox}%
    \sbox\algocf@inoutbox{\parbox[t]{\inoutsize}{\KwSty{#2}\algocf@typo:\hfill}~}\setlength{\inoutindent}{\wd\algocf@inoutbox}%
    \fi%
  \fi
  \algocf@newcommand{#1}[1]{%
    \ifthenelse{\boolean{algocf@inoutnumbered}}{\relax}{\everypar={\relax}}%
    {\let\\\algocf@newinout\hangindent=\inoutindent\hangafter=1\parbox[t]{\inoutsize}{\KwSty{#2}\algocf@typo:\hfill}~##1\par}%
    \algocf@linesnumbered
  }}%
\makeatother

\begin{document}

\title{CertainSync: Rateless Set Reconciliation with Certainty}


\author{Tomer Keniagin}
\email{tkeniagin@campus.technion.ac.il}
\orcid{0009-0004-2711-104X}
\affiliation{%
   \institution{Technion - Israel Institute of Technology}
   \city{Haifa}   \country{Israel}}

\author{Eitan Yaakobi}
\email{yaakobi@cs.technion.ac.il}
\orcid{0000-0002-9851-5234}
\affiliation{%
   \institution{Technion - Israel Institute of Technology}
   \city{Haifa}   \country{Israel}}

\author{Ori Rottenstreich}
\email{or@technion.ac.il}
\orcid{0000-0002-4064-1238}
\affiliation{%
   \institution{Technion - Israel Institute of Technology}
   \city{Haifa}   \country{Israel}}


\begin{abstract}
Set reconciliation is a fundamental task in distributed systems, particularly in blockchain networks, where it enables synchronization of transaction pools among peers and facilitates block dissemination. Traditional set reconciliation schemes are either statistical, offering success probability as a function of communication overhead and symmetric difference size, or require parametrization and estimation of that size, which can be error-prone. We present \emph{CertainSync}, a novel reconciliation framework that, to the best of our knowledge, is the first to guarantee successful set reconciliation without any parametrization or estimators. The framework is rateless and adapts to the unknown symmetric difference size. Reconciliation is guaranteed whenever the communication overhead reaches a lower bound derived from the symmetric difference size and universe size. Our framework builds on recent constructions of Invertible Bloom Lookup Tables (IBLTs), ensuring successful element listing as long as the number of elements is bounded. We provide a theoretical analysis proving the certainty of reconciliation for multiple constructions. Our approach is validated by simulations, showing the ability to synchronize sets with efficient communication costs while maintaining guarantees compared to baseline schemes. To further reduce overhead in large universes such as blockchain networks, CertainSync is extended with a universe reduction technique. We compare and validate this extension, \emph{UniverseReduceSync}, against the basic framework using real Ethereum transaction hash data. Results show a trade-off between lower communication costs and maintaining guarantees, offering a comprehensive solution for diverse reconciliation scenarios.
\end{abstract}





\maketitle



\section{Introduction}
Set reconciliation is essential for synchronizing data across systems like cloud storage services \cite{CLOUD_STORAGE_SERVICES}, Peer-to-Peer (P2P) networks~\cite{SET_RECONCILIATION_IN_P2P}, distributed computing~\cite{MULTI_PARTY_SET_RECONCILIATION}, and blockchain networks~\cite{GRAPHENE}. Instead of directly exchanging set elements, which incurs high communication overhead $\mathcal{O}(|A| + |B|)$ for sets $A$ and $B$, reconciliation protocols use compact representations based on coding theory and probabilistic data structures (sketches). These representations enable efficient identification of symmetric differences with lower communication costs. For instance, in blockchain networks, efficient reconciliation enables light clients to synchronize their transaction pools and blockchain states effectively, as seen in protocols like MempoolSync~\cite{MEMPOOLSYNC} and SREP~\cite{SREP}. Additional reconciliation applications in fields like collaborative editing and distributed databases are detailed in Appendix~\ref{appx:applications}.

\begin{figure}[h!] 
\centering
\includegraphics[width=0.80\linewidth]{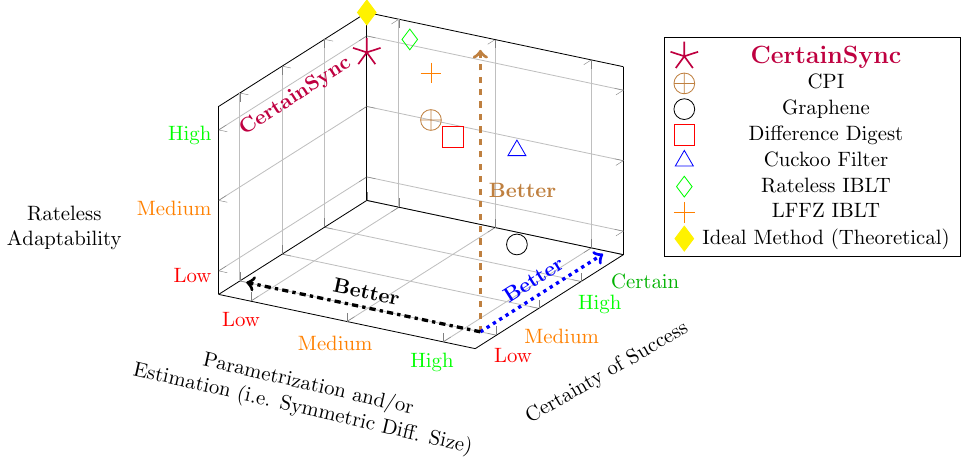}
\caption{Comparison of set reconciliation schemes based on three metrics: (i) parametrization tuning and/or symmetric difference size estimation overhead (lower is better), (ii) certainty of success (higher is better), and (iii) rateless adaptability (higher is better) indicates efficiency by dynamically adapting to varying set differences while minimizing communication overhead and retransmissions. A detailed explanation of the parameters and estimators used for each scheme is presented in Table \ref{tab:parameterization} in Appendix \ref{appx:set_reconciliation_solutions}.}
\label{fig:methods_comparison_intro}
\end{figure}

Moreover, data consistency (via guaranteed identification of symmetric differences), as in blockchain systems, is crucial.
Reconciliation schemes fall into two main categories: probabilistic and exact. Probabilistic approaches, such as Rateless IBLT~\cite{RATELESS_IBLT} or with Cuckoo filter \cite{SET_RECONCILIATION_WITH_CUCKOO_FILTER, mcfsyn}, offer low overhead but risk occasional failures in identifying symmetric differences. Exact schemes, like CPISync~\cite{CPISync} and PinSketch~\cite{PinSketch}, guarantee the success of identifying symmetric differences by relying on symmetric difference size estimation or an upper bound, which can be computationally expensive and prone to errors. Additionally, existing reconciliation schemes often involve complex parameter tuning as in Graphene \cite{GRAPHENE} for set reconciliation among peers in blockchains and related distributed systems, where there is a parameter search algorithm based on input parameters given, or estimations such as Strata Estimator in Difference Digest \cite{EFFICIENT_SET_RECONCILIATION} for set difference size estimation. In large-scale or real-time set reconciliation applications, where simplicity, scalability, and low latency are prioritized, parametrization and estimators, which incur significant overhead and complexity, should be avoided.
This work introduces a novel reconciliation framework called CertainSync that combines two desirable properties: \emph{(i)} simplicity, with no parameter tuning or difference size estimation, and \emph{(ii)} guaranteed success in computing symmetric differences through rateless transmission. Fig.~\ref{fig:methods_comparison_intro} illustrates how our approach achieves these goals, ensuring both robustness and efficiency. 
Our focus is on guaranteeing successful set reconciliation in a parameterless manner, making this the first work to introduce this concept.

CertainSync can be tuned with one of three proposed constructions based on which elements are mapped to the IBLTs. Table~\ref{table_constructions} summarizes the basic properties of the three constructions as a function of the following parameters: (i) universe size from which elements can appear ($n$), and (ii) size of the symmetric difference ($|\Delta|$). Each construction is characterized by 
(i) the maximum supported symmetric difference size ($|\Delta|_{\text{max}}$), 
(ii) the incremental communication overhead per single element increase in the symmetric difference size ($\Delta m_d$), 
and 
(iii) the total communication overhead ($m_d$). 
In Table~\ref{table_constructions}, the total communication overhead refers to the number of IBLT cells required for synchronization. Throughout the paper, the comparison of set reconciliation schemes is measured in bits, derived from the respective bit size of an individual
IBLT cell, with additional overhead bits as necessitated by the distinct requirements of each scheme.
To evaluate the characteristics of our proposed framework, we conduct experiments using datasets that include sets of positive numbers to mimic the transaction pools in blockchain systems. Our experimental results demonstrate that our framework successfully computes the symmetric difference with certainty, and communication overhead is determined by a bound derived from the symmetric difference size and the universe size. 
Since  this bound depends on the universe size, which can be large, it may result in inefficiencies in large-scale settings like blockchain networks. We mitigate this issue by extending our framework to include an optional universe size reduction.
We provide a comprehensive comparative analysis of existing set reconciliation schemes compared to our proposed framework. Also, we apply our framework in the Ethereum blockchain with real blockchain data to synchronize transaction pools. 
The notations used throughout the paper are presented in Table \ref{tab:notations} in Appendix \ref{appx:additional_tables}.
\begin{table*}[t!]
\centering
\caption{Proposed constructions for CertainSync and their inherent properties. 
$d$ represents the  symmetric difference size ($|\Delta|$) and $n$ indicates the universe size.}
\resizebox{\textwidth}{!}{\begin{tabular}{|c|c|c|c|}
\hline
\makecell{CertainSync\\ Construction} & \makecell{Max Symmetric \\ Difference Size \\ $|\Delta|_{max}$} & \makecell{\hfill \\ Incremental Communication  \\ Overhead \\ $\Delta m_d$ (Eq. (\ref{cells_difference})) \\ \hfill} & \makecell{Total Communication \\ Overhead \\ $m_d$} \\
\hline 
\makecell{Construction \RNum{1} \\ EGH \cite{EGH} \\ (Subsection \ref{egh_construction})} & $n$ & \makecell{ \\ 
$ O\del{\frac{d\log{n}}{\log{d}+\log{\log{n}}}}$} & \makecell{ \\
$ O\left(\frac{d^2 \log^2 n}{\log d + \log \log n}\right) $} \\
\hline
\makecell{Construction \RNum{2} \\ OLS \cite{OLS} \\ (Subsection \ref{ols_construction})} & $\ceil{\sqrt{n}}$ & $\ceil{\sqrt{n}}$ & $d\ceil{\sqrt{n}}$ \\ 
\hline 
\makecell{Construction \RNum{3} \\Extended Hamming \cite{EXTENDED_HAMMING_CODE}\\ (Subsection \ref{extended_hamming_construction})} & $3$  &  $\lceil \log_2 n \rceil$ & $(d-1) \lceil \log_2 n \rceil + 1$ \\
\hline
\end{tabular}}
\label{table_constructions}
\end{table*}
We make available the implementation of CertainSync developed in Python and Go as an open source on GitHub\footnote{\url{https://github.com/toto9820/Rateless-Set-Reconciliation-with-Listing-Guarantees}}. It includes algorithms implementations,  datasets and test scripts for reproducibility.

\section{Background and Related Work}

\subsection{Set Reconciliation}
In set reconciliation, two or more parties hold finite sets 
and aim to learn the elements missing from each other's sets. 
Set reconciliation often utilizes the symmetric difference operation to efficiently identify and reconcile discrepancies between sets. Key performance metrics for evaluating set reconciliation schemes include: \emph{(i)} decoding accuracy (probability of correctly identifying the symmetric difference),  \emph{(ii)} communication efficiency (minimizing total communication overhead),  \emph{(iii)} scalability (minimizing additional communication overhead as 
symmetric difference size increases), \emph{(iv)} computation overhead (processing requirements at each party), and \emph{(v)} total reconciliation time (the time required until the sets are fully reconciled) are of great interest. However, the primary focus is on decoding accuracy and communication overhead, as less system-independent factors (e.g. network bandwidth, compute capabilities), offering a more general basis for evaluation of different set reconciliation schemes.
It is noteworthy that computing the union of sets is a fundamental operation that can be executed by any participant involved in the set reconciliation process. We denote by $\Delta$ the symmetric difference of two considered sets. 

\begin{definition}[Symmetric Difference]
The symmetric difference for two sets $A$ and $B$ 
refers to elements that appear in one set and not in the other set, namely 
$\Delta = (A \setminus B) \cup (B \setminus A)$. 
\end{definition} 
For example, given two sets \( A = \{1, 2, 3, 4\} \) and \( B = \{3, 4, 6, 7\} \), the symmetric difference consists of elements in \( A \) but not in \( B \) (\(A \setminus B = \{1, 2\} \)) and elements in \( B \) but not in \( A \) (\(B \setminus A = \{6, 7\} \)). Thus, the symmetric difference is \( \Delta = (A \setminus B) \cup (B \setminus A) = \{1, 2, 6, 7\} \).

\subsection{Invertible Bloom Lookup Table (IBLT) and its Applications}
Sketches, such as Bloom Filter (BF) \cite{BF} and Count-Min \cite{COUNT_MIN_SKETCH}, provide space-efficient, probabilistic representations of sets. 
These methods trade off some accuracy for substantial gains in efficiency and scalability, making them suitable for applications where approximate results are acceptable.
An IBLT \cite{IBLT} is another sketch type that combines efficiency with the ability to identify the elements inserted into it probabilistically. An IBLT consists of a fixed-size array of cells, where each cell contains the following fields: \emph{(i)} count, an integer representing the number of elements mapped to this cell, \emph{(ii)} xorSum, the XOR of all elements mapped to this cell, and \emph{(iii)} checkSum, the XOR of the hash values of all elements mapped to this cell. The operations supported by an IBLT are insertion for adding an element to an IBLT, deletion for removing an element from an IBLT, and listing for retrieving elements from an IBLT. 
An important concept in IBLT listing is a \emph{pure cell}. This is a cell that contains a single inserted element. A pure cell is identifiable by containing only one element mapped to it.  
The listing procedure repeatedly searches for a pure cell and identifies the element based on the xorSum field when found. The identified element is then removed from other cells it is mapped to, reducing the remaining elements in those cells. The process fails if no pure cells are found before listing all elements.

In this paper, we mainly focus on set reconciliation approaches based on IBLTs.  
IBLTs offer several advantages for set reconciliation: they provide a compact representation of sets, allowing for minimal communication of set differences \cite{EFFICIENT_SET_RECONCILIATION}, they support the insertion and deletion of elements, and there is support for listing elements with success probability. By exchanging IBLT sketches, parties can reconcile their sets efficiently.  
IBLTs could represent elements that are present in one set but absent in the other, as demonstrated in Difference Digest \cite{EFFICIENT_SET_RECONCILIATION} by the subtraction operation.
However, they have a major drawback - the listing is not guaranteed to succeed as there is a success probability derived from the number of elements inserted into the IBLT and the IBLT memory size represented as its number of cells.
The listing success probability of an IBLT, as shown in \cite{IBLT}, is almost 1 ($1-o(1)$) if the ratio between the number of available cells and the number of inserted elements exceeds a certain threshold $c_h$, where $h$ is the number of hash values. In contrast, if the ratio falls below \( c_h \), as described in \cite{IBLT_LISTING_PROBABILITY}, partial listing (namely partial extraction representing listing success probability less than 1) becomes more likely. However, a small number of iterative executions of partial
listing can eventually achieve a full listing of all elements inserted in the IBLT.

\subsection{Concepts in Coding Theory and Bloom Filters}

For any positive integer \( k \), we define \( [k] \coloneqq \{1, \ldots, k\} \) and \( [k]_0 \coloneqq \{0, \ldots, k-1\} \), and these notations are used consistently throughout the paper to represent sets.

\subsubsection{Bloom Filter with FPFZ}
Bloom filter with a FPFZ (False Positive Free
Zone) \cite{BLOOM_FILTER_WITH_EGH} is a specific Bloom filter construction where for a set of up to $d$ elements from a finite universe $U$, membership queries return true if and only if an element is actually in the set, thus guaranteeing no false positives and no false negatives for any membership queries.

\subsubsection{EGH \& EGH Bloom Filter}
The Eppstein, Goodrich, and Hirschberg (EGH) method \cite{EGH} was originally developed for group testing by leveraging properties of modular arithmetic and the Chinese Remainder Theorem. However, this method has been adapted for various applications, including Bloom filters \cite{BLOOM_FILTER_WITH_EGH} with EGH filter. In the context of Bloom filters, the EGH method establishes an upper bound $d$ on the maximum number of inserted elements for which a FPFZ is guaranteed. This upper bound is related to the size of the universe $n$ and a product of prime numbers, expressed as $d \leq \log_n \Pi_k = \sum_{j=1}^{k} \log_n p_j$, where $\Pi_k \triangleq \prod_{j=1}^{k} p_j$ is the product of the first $k$ prime numbers, and $p_j$ represents the $j$-th prime number ($p_1 = 2, p_2 = 3, p_3 = 5, \ldots$). 

\subsubsection{OLS \& OLS Bloom Filter}
Orthogonal Latin Square (OLS) codes \cite{OLS_CODES} have been widely
studied to protect memories from errors, as they have a modular construction and can be decoded in parallel with simple
circuitry \cite{EXTEND_OLS_CODES}.
A Latin square of order $n$ is an $n \times n$ array filled with $n$ different symbols where each symbol belongs to the set $[n]_0$, and each symbol occurs exactly once in each row and exactly once in each column \cite{LATIN_SQUARES_BOOK} as shown in Fig. \ref{ols_code_examples}(a). 
Two Latin squares are orthogonal if, when superimposed, each ordered pair of symbols appears exactly once as shown in Fig. \ref{ols_code_examples}(b). 
A set of Latin squares of the same order such that every pair of squares is orthogonal is called Mutually Orthogonal Latin Squares (MOLS); it is denoted by \(\text{MOLS}(n)\) where $n$ is the order of the Latin squares. 
The maximum size of \(\text{MOLS}(n)\), as shown in \cite{LATIN_SQUARES_BOOK} (Theorem 5.1.2), is at most $n-1$.
A set of $n-1$  \(\text{MOLS}(n)\) is called a complete set of MOLS.
When $n$ is a prime power, we are guaranteed to have at least one complete set of \(\text{MOLS}(n)\), as stated in \cite{LATIN_SQUARES_BOOK} (Theorem 5.2.3).
OLS filter is a Bloom filter constructed
using OLS code \cite{OLS}. Besides using MOLS, two additional special matrices (which are not
Latin squares) are used for the construction of the filter. 
One of the two additional special matrices used later we notate as $R_n$, where each of its rows consists of a single repeated value, ranging from 0 to $n-1$. Specifically, the $i$-th row of $R_n$ matrix is filled with the value $i-1$ for all columns, as illustrated in Fig. \ref{ols_code_examples}(c). The second special matrix is its transpose $R_n^{T}$.
\begin{figure}[h!]
\begin{minipage}[t]{0.3\textwidth}
\centering
\begin{tabular}{ccccc|ccccc}
\hline
0 & 1 & 2 & 3 & 4 & 0 & 1 & 2 & 3 & 4 \\
1 & 2 & 3 & 4 & 0 & 2 & 3 & 4 & 0 & 1 \\
2 & 3 & 4 & 0 & 1 & 4 & 0 & 1 & 2 & 3 \\
3 & 4 & 0 & 1 & 2 & 1 & 2 & 3 & 4 & 0 \\
4 & 0 & 1 & 2 & 3 & 3 & 4 & 0 & 1 & 2 \\
\end{tabular}
\caption*{(a) Two Latin squares of order 5.}
\label{fig:latin_squares}
\end{minipage}
\hspace{1.0cm}
\begin{minipage}[t]{0.3\textwidth}
\centering
\begin{tabular}{ccccc}
0,0 & 1,1 & 2,2 & 3,3 & 4,4 \\
1,2 & 2,3 & 3,4 & 4,0 & 0,1 \\
2,4 & 3,0 & 4,1 & 0,2 & 1,3 \\
3,1 & 4,2 & 0,3 & 1,4 & 2,0 \\
4,3 & 0,4 & 1,0 & 2,1 & 3,2
\end{tabular}
\caption*{(b) Superposition of  two Latin squares.}
\label{fig:superposition}
\end{minipage}
\hfill
\begin{minipage}[t]{0.3\textwidth}
\centering
\begin{tabular}{ccccc}
0 & 0 & 0 & 0 & 0 \\
1 & 1 & 1 & 1 & 1 \\
2 & 2 & 2 & 2 & 2 \\
3 & 3 & 3 & 3 & 3 \\
4 & 4 & 4 & 4 & 4
\end{tabular}
\caption*{(c) Additional matrix $R_5$ for OLS construction.}
\label{fig:additional_matrice}
\end{minipage}
\caption{Latin squares, superposition, and additional matrix for constructing OLS code.}
\label{ols_code_examples}
\end{figure}
\subsubsection{Extended Hamming Code}
An Extended Hamming code is an Error Correcting Code (ECC) that extends the standard Hamming code by adding an extra parity bit to have a 1-bit error-correcting or 3-bit error-detecting code.

\begin{definition}[Extended Hamming Code]
An Extended Hamming code is a linear code denoted by $(n=2^m, k=2^m - m - 1, d_H=4)$, where $n$ is the codeword length; $k$ is the information length (the dimension); $d_H$ is the minimum Hamming distance between any two codewords, and $m\geq 2$ is a positive integer. The parity check matrix $H_n$ of the Extended Hamming code is the following:
\[
H_n = 
\begin{bmatrix}
    1 & 1 & 1 & \cdots & 1 & 1 \\
    & & H_n^{'}
\end{bmatrix}, 
\quad
H_n^{'} = 
\begin{bmatrix}
    0 & 0 & 0 & \cdots & 1 & 1 \\
    \vdots & \vdots & \vdots & \ddots & \vdots & \vdots \\
    0 & 0 & 1 & \cdots & 1 & 1 \\
    0 & 1 & 0 & \cdots & 0 & 1 \\
\end{bmatrix}.
\]
where the matrix $H_n^{'}$ consists of all the \( 2^m \) binary column vectors of length \( m = \log_2(n)\). 
\end{definition}

\subsubsection{Stopping Set}
A stopping set \cite{STOPPING_SET} is a combinatorial structure, originally defined in the context of matrices for the decoding procedure of error-correcting codes \cite{STOPPING_SET_ECC}.

\begin{definition}[Stopping Set]
Let $M$ be a matrix and $S$ a non-empty set of its columns.
The row's weight in the sub-matrix implied by $S$ is defined as the number of non-zero coordinates in it. $S$ is called a stopping set if it has no row of weight one. The stopping distance of $M$, denoted by $s(M)$, is the size of the smallest stopping set in $M$.   
\end{definition}

\subsection{Blockchain Application of Set Reconciliation}
\begin{figure}[h!] 
\centering
\includegraphics[width=0.45\textwidth]{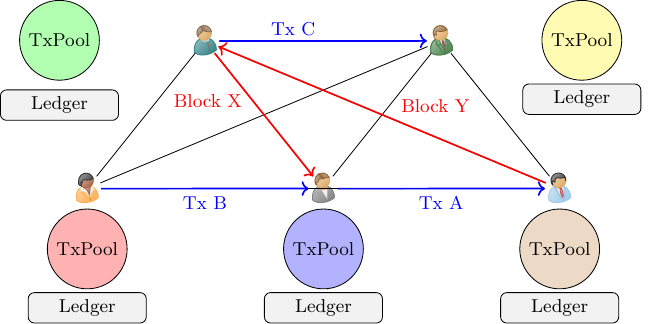}
\caption{Blockchain network modeled as a peer-to-peer (P2P) system where each peer maintains a local ledger (immutable chain of blocks) and a transaction pool (TxPool). Peers synchronize transactions and mined blocks to achieve consensus across the network.}
\label{fig:blockchain_peers_scheme}
\end{figure}

Set reconciliation plays a crucial role in various applications, and one prominent application is in blockchain networks.
Blockchain networks rely on efficient and reliable synchronization of transactions from transaction pools (TxPools) and newly mined blocks (block
propagation process) between peers, as illustrated in Fig. \ref{fig:blockchain_peers_scheme} to maintain consensus and ensure the integrity for proper  functioning. 
IBLTs play an important role in blockchain networks by enabling efficient representation and reconciliation of transaction pools and blocks between peers,  significantly reducing communication overhead as demonstrated in Graphene \cite{GRAPHENE} for interactive set reconciliation in blockchain networks. 
Various
protocols have been proposed for the synchronization of transaction pools outside (and independently) of the block propagation channel \cite{MEMPOOLSYNC, SREP, PRESYNC} due to long block transmission time and block validation time like SREP \cite{SREP}, which leverages out-of-band synchronization of transaction pools, ensuring that only differences between transaction pools are exchanged to minimize communication overhead. 

\subsection{Rateless Coding \& IBLT}
Rateless coding, also known as fountain coding, enables efficient and reliable data transmission over unreliable or lossy channels. Unlike traditional fixed-rate coding schemes, rateless codes do not have a predetermined rate or block length. Instead, an unlimited number of encoded symbols are generated from the original data, allowing the recipient to recover the original data by collecting a sufficient number of these encoded symbols, regardless of which specific symbols are received. 
\begin{definition}[Rateless Coding]
\label{Rateless_Coding}
Let $\mathcal{S}$ be the original data of size $k$ symbols, and $\mathcal{C} = \{c_1, c_2, \ldots, c_n\}$ be the set of encoded symbols generated by the rateless code. Recipient can reconstruct $\mathcal{S}$ from any subset $\mathcal{C}' \subseteq \mathcal{C}$ such that $|\mathcal{C}'| \geq t(k)$, where $t(k)$ is a threshold function dependent on the size $k$ of the original data and determined by the specific coding scheme.
\end{definition}
Rateless coding has been recently applied to the problem of set reconciliation~\cite{RATELESS_IBLT}. In this work, the authors propose a rateless set reconciliation protocol based on the original IBLT approach with hash functions. Instead of encoding a set into a fixed-size IBLT, the rateless IBLT approach generates an unbounded stream of IBLT cells as coded symbols. Participant 1 can continuously transmit these coded symbols until Participant 2 has collected enough IBLT cells to decode the set difference successfully.   
In Definition~\ref{Rateless_Coding}, $k$ represents the symmetric difference size $|\Delta|$. Also, Lázaro et al. \cite{RATELESS_MET_IBLT} proposed a rateless solution based on a variant of IBLT named MET IBLT.

\subsection{Listing Failure Free Zone (LFFZ) IBLT}
One significant challenge in set reconciliation is ensuring that all the set elements are correctly identified, known as the listing guarantee. Traditional methods often fail to provide this guarantee due to their inherent false positive rate. Schemes to provide this listing guarantee have been proposed in the past:
Bloom filters with a trie-based mechanism for eliminating false positives \cite{Extended_BF_WITH_TRIE}, changing the traditional decoding method with pure cells by asserting an extra pure cell condition \cite{DECODING_ERRORS_DIFF_IBLT}, an additional stash data structure based on error correcting codes, which serves as a backup when IBLT fails to decode correctly \cite{SPACE_TIME_TRADE_OFFS_FOR_ROBUST_SET_RECONCILIATION}, and replacing the random hash functions for mapping elements of the set to IBLT cells, which contribute to the probabilistic nature of the IBLT, with multiple constructions that are based on various coding techniques \cite{LFFZ_IBLT}. 
The concept of IBLTs with a Listing Failure Free Zone (LFFZ)~\cite{LFFZ_IBLT}, provides a guarantee of successful listing for all sets up to a certain size parameter $d$, thereby enhancing the reliability and robustness of IBLTs.
An $(n,d)$-LFFZ IBLT is defined as follows:

\begin{definition}[\LFFZ{d}]
\label{def:lffz}
Let $U = [n]$ be a finite universe of size $n$, and let $S \subseteq U$ be a set of size at most $d$. If an (n, d)-LFFZ IBLT is constructed to encode the set $S$, then the decoding process is guaranteed to successfully list all elements in $S$, regardless of the specific elements or their distribution within the universe $U$.
\end{definition}

LFFZ IBLTs rely on several combinatorial and recursive methods for their construction. They use a binary mapping matrix $M$ of size $m \times n$ to map elements of a set to cells in an IBLT, instead of using hash functions as in the original IBLT \cite{IBLT}. 

\begin{definition}[Binary Mapping Matrix $M$]
A binary mapping matrix \( M \) of size \( m \times n \) is used to map elements of a set to cells in an IBLT. Each row of the matrix corresponds to a cell of an IBLT, and each column corresponds to an element in the universe. The entry $M[i][j]$ indicates whether the $j$-th element is mapped to the $i$-th cell, with $M[i][j] = 1$ if the element is mapped to the cell, and $M[i][j] = 0$ otherwise.    
\end{definition}

An important concept from~\cite{LFFZ_IBLT} assures that for any set $S$ of at most $d$ elements from $U$, the mapping matrix $M$ does not contain any stopping set of size at most $d$. This leads to the definition of a \decodable{} matrix.

\begin{definition}[\decodable{} matrix]\label{def:d-decodable}
An $m \times n$ binary matrix $M$ is called \decodable{} if its stopping distance is at least $d+1$, that is, $s(M) \geq d+1$. Given $n, d \in \mathbb{N}$, with $d \leq n$, the minimal number of rows of a \decodable{} matrix is denoted by $m^*(n,d)$, that is,
\[
m^*(n,d) = \min \cbr{m : \exists M \in \cbr{0,1}^{m \times n}, M \text{ is \decodable{}}}.
\]
\end{definition}

From the definition of a $d$-decodable matrix, it follows that any \decodable{} matrix is also a $(d-1)$-decodable matrix.

\section{Set Reconciliation Using the C\lowercase{ertain}S\lowercase{ync} Framework}

\label{sec:certain_sync_framework}

In the CertainSync framework, we utilize the constructions of \( d \)-decodable rateless matrices (where \( d = |\Delta| \)) from Section~\ref{section:constructions} as mapping matrices. These matrices map elements to IBLT cells, enabling set reconciliation with certainty. The certainty guarantees the listing success in retrieving the symmetric difference $\Delta$ under certain conditions without knowing the size of the symmetric difference $|\Delta|$ in advance. 

\subsection{Two-Party Problem for the CertainSync Framework}
Let $U = [n]$ be a finite universe of size $n$, and let $S_1, S_2 \subseteq U$ be the sets held by Participant~1 ($P_1$) and Participant~2 ($P_2$), respectively. Given a $d$-decodable rateless matrix $M_{n,d}$ of size $m \times n$, where $m$, which depends on the unknown symmetric difference size, is the finite 
number of cells in an IBLT. $P_1$ constructs $\text{IBLT}_1$ from set $S_1$ according to the mapping defined by $M_{n,d}$, and then sends $\text{IBLT}_1$ cells to $P_2$ in a rateless manner. The objective is to construct an $\text{IBLT}$ of the symmetric difference ($\text{IBLT}\{\Delta\}$) at $P_2$ by performing the subtraction of the IBLTs of $P_2$ and $P_1$, with the guarantee that the listing operation at $P_2$ successfully recovers all elements in $\Delta$. Upon successful recovery of $\Delta$, $P_2$ could transmit the required subset $S_2 \setminus S_1$ of $\Delta$  to $P_1$ to achieve complete set reconciliation. 

\subsection{Set Reconciliation with Certainty}
A set reconciliation with certainty is a more stringent variant of exact set reconciliation, where the goal is not only to fully recover the symmetric difference between two sets, but also to do so  without prior knowledge of the size of the symmetric difference, defined formally as follows: 

\begin{definition}[Set Reconciliation with Certainty]
Given two participants $P_1$ and $P_2$ holding sets $S_1, S_2 \subseteq U$ respectively, where $U = [n]$ is a finite universe of size $n$. Set reconciliation with certainty is achieved if 
it correctly recovers the symmetric difference $\Delta = (S_1 \setminus S_2) \cup (S_2 \setminus S_1)$ with probability 1, succeeds in recovery without requiring prior knowledge of the symmetric difference size $|\Delta|$, and terminates after a finite number of steps.
\end{definition}

\begin{figure}[h]
    \centering
    \begin{minipage}[t]{0.45\textwidth}
        \centering
        \includegraphics[width=\textwidth]{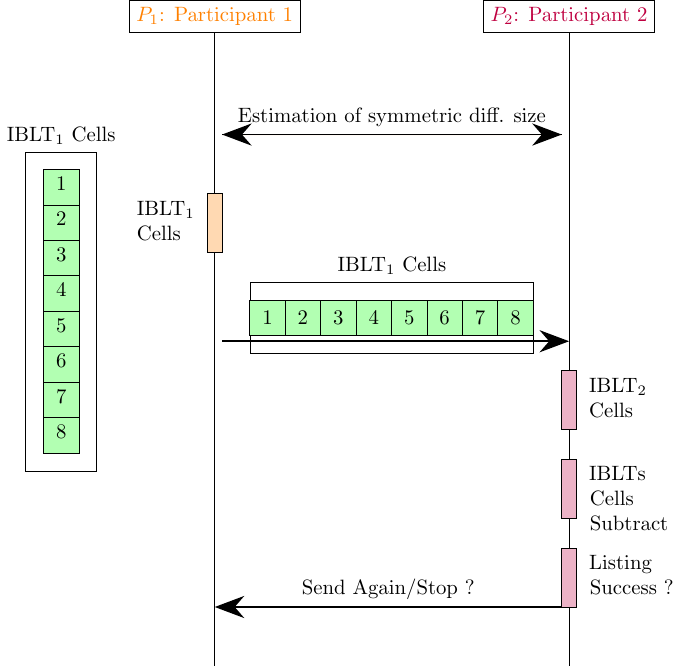}
        
        \caption*{(a) Traditional Set Reconciliation}
        \label{fig:traditional_set_reconciliation}
    \end{minipage}
    \hfill
    \begin{minipage}[t]{0.45\textwidth}
        \centering
        \includegraphics[width=\textwidth]{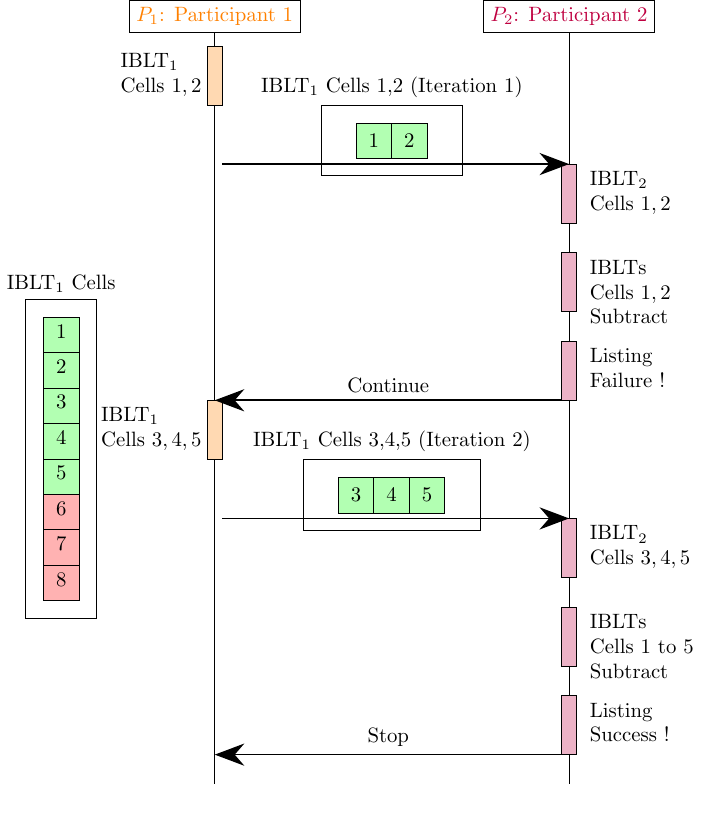}
        
        \caption*{(b) Set Reconciliation with CertainSync}
        \label{fig:our_set_reconciliation}
    \end{minipage}
    \caption{Illustration of traditional set reconciliation vs. CertainSync.}
    \label{fig:set_reconciliation_comparison}
\end{figure}

In the CertainSync framework, set reconciliation with certainty is achieved by leveraging \( d \)-decodable rateless matrices, which enable set reconciliation without prior knowledge of the symmetric difference size and termination in a finite number of steps.
Fig. \ref{fig:set_reconciliation_comparison}(a) illustrates the traditional set reconciliation as described in Algorithm \ref{alg:Traditional_SetReco} in Appendix \ref{appx:additional_algorithms}, showing how the IBLTs of the participants are constructed with a fixed number of cells based on estimation and/or parametrization, and the probabilistic listing of their subtraction to determine the symmetric difference between them with success probability.
In our set reconciliation framework, as described in Fig. \ref{fig:set_reconciliation_comparison}(b), using CertainSync, Participant 1 constructs at each iteration an amount (according to chosen construction) of $\text{IBLT}_1$ cells representing its set and transmits it to Participant 2. There is no estimation of the symmetric difference size at any point like in the traditional set reconciliation.
At each iteration, Participant 2 constructs the same amount of cells received from Participant 1, and checks if they have enough total cells to ensure the success of the listing of $\text{IBLT}\{\Delta\}$ to find the symmetric difference between the two sets with certainty (success probability of 1).
If so, Participant 2 tells Participant 1 to stop sending cells, and lists successfully all the elements in $\text{IBLT}\{\Delta\}$ to find the symmetric difference between the two sets with certainty.

\subsection{Applications for CertainSync Constructions}

Constructions of CertainSync and their extended variants with universe reduction (UniverseReduceSync, presented in Section~\ref{sec:universe_reduce_sync_framework}) are tailored to address various scenarios detailed in Table~\ref{tab:certainsync_applications}, therefore offering a comprehensive solution for set reconciliation with certainty.
\begin{table}[h!]
\centering
\caption{Comparison of proposed CertainSync constructions and their practical applications.}
\resizebox{\textwidth}{!}{\begin{tabular}{|c|c|c|c|}
\hline
\backslashbox{Property}{\makecell{CertainSync \\ Construction}} & \makecell{Construction \RNum{1} \\ EGH \cite{EGH} \\ (Subsection \ref{egh_construction})} & \makecell{Construction \RNum{2} \\ OLS \cite{OLS} \\ (Subsection \ref{ols_construction})} & \makecell{Construction \RNum{3} \\Extended Hamming \cite{EXTENDED_HAMMING_CODE}\\ (Subsection \ref{extended_hamming_construction})} \\
\hline
Purpose & 
\begin{tabular}[c]{@{}c@{}}Unbounded Symmetric \\ Difference Size
\end{tabular} & 
\begin{tabular}[c]{@{}c@{}}Medium Symmetric Difference Size \\ \& Large Bounded Universe\end{tabular} & 
\begin{tabular}[c]{@{}c@{}}Small Symmetric \\ Difference Size \end{tabular} \\
\hline
Key Advantage & 
\begin{tabular}[c]{@{}c@{}}Flexible  Symmetric \\ Difference\end{tabular} & 
\begin{tabular}[c]{@{}c@{}}Improved Communication \\ Overhead Scalability\end{tabular} & 
\begin{tabular}[c]{@{}c@{}}Low Communication \\ Overhead\end{tabular} \\
\hline
Small Universe Compatible & \faCheckSquareO \hspace{0.01cm} CertainSync & \faCheckSquareO \hspace{0.01cm} CertainSync & \faCheckSquareO \hspace{0.01cm} CertainSync \\
\hline
Medium Universe Compatible & \faCheckSquareO \hspace{0.01cm} CertainSync & \faCheckSquareO \hspace{0.01cm} CertainSync & \\
\hline
Large Universe Compatible & \makecell{\faCheckSquareO \hspace{0.01cm} CertainSync \\ \faCheckSquareO \hspace{0.01cm} UniverseReduceSync } & \faCheckSquareO \hspace{0.01cm} UniverseReduceSync & \\
\hline
Potential Applications & 
\begin{tabular}[c]{@{}l@{}}
- Blockchain synchronization \\ 
- Highly dynamic data systems \\ 
\end{tabular} & 
\begin{tabular}[c]{@{}l@{}}
- Collaborative document editing \\ 
- Distributed database sync \\ 
- Version control systems
\end{tabular} & 
\begin{tabular}[c]{@{}l@{}}
- Error correction 
\end{tabular} \\
\hline
\end{tabular}}
\label{tab:certainsync_applications}
\end{table}
Intuitively, CertainSync EGH is well-suited for blockchain synchronization due to its ability to handle varying symmetric differences over time in a full universe-size range. CertainSync OLS is useful for the synchronization of distributed databases or files with an overall fixed size (fixed universe size). Meanwhile, CertainSync Extended Hamming is relevant for error correction, offering a low communication overhead solution for managing small symmetric differences effectively.

\subsection{Algorithms for CertainSync Framework}

We present an overview of the key algorithms for implementing and utilizing CertainSync. These algorithms are designed to ensure guaranteed listing success and utilize rateless adaptability. The detailed algorithms are provided in Appendix \ref{appx:additional_algorithms} with an example in Appendix \ref{appx:additional_examples}.

\textbf{ConstructIBLT.} This algorithm constructs at each iteration $i$ an amount of IBLT cells (according to chosen construction) from a given set $S$ using submatrix of mapping matrix $M_{n,d}$, denoted as $M_{n,i}$, which is an $i$-decodable rateless matrix. The IBLT is initialized with cells containing count, xorSum, and checkSum fields, which are updated as elements from the set are processed.

\textbf{IBLTDiff.} This algorithm constructs the IBLT of the symmetric difference ($\text{IBLT}\{\Delta\}$) by subtracting the IBLT of $P_1$ ($\text{IBLT}_1$) from the IBLT of $P_2$ ($\text{IBLT}_2$).

\textbf{DecodeDiff.} This algorithm lists the symmetric difference $\Delta$ by decoding the IBLT of the symmetric difference ($\text{IBLT}\{\Delta\}$). It retrieves elements from pure cells and removes them from other IBLT cells until all elements are decoded or a failure occurs if the IBLT is not empty at the end. 


\section{Constructions for C\lowercase{ertain}S\lowercase{ync} Framework}
\label{section:constructions}
For our constructions, we use a subfamily of \decodable{} matrices that inherently possess a rateless structure, defined as follows:

\begin{definition}[\decodable{} Rateless Matrix]
\label{def:decodable_rateless}
An \( m \times n \) matrix \( M_{n,d} \) is called a \( d \)-decodable rateless matrix if there exist positive integers \( m_1 \leq m_2 \leq \cdots \leq m_d = m \), such that for every \( i \in [d] \), the \( m_i \times n \) submatrix formed by the first \( m_i \) rows of \( M_{n,d} \) is \( i \)-decodable. We refer to the vector \( (m_1, m_2, \ldots, m_d ) \)  as the \textit{decodability profile} of the matrix \( M_{n,d} \).
\end{definition}

The decodability profile of a \decodable{} rateless matrix indicates the additional number of rows required to transition from \((i-1)\)-decodability to \(i\)-decodability, where $i \in [d]$, capturing the incremental growth in rows needed for progressively increasing decodability. Specifically, let \( \Delta m_i \) denote the additional number of rows added when transitioning from \( (i-1) \)-decodable to \( i \)-decodable. With \(\Delta m_1 = m_1\) i.e. $m_0 = 0$ the additional number of rows added \( \Delta m_i \) is the following:
\begin{equation}
\label{cells_difference}
 \Delta m_i = m_i - m_{i-1},
\end{equation}

\textbf{Example (\(d\)-Decodable Rateless Matrix).}
For \(n = 8\) columns, we construct a \(2\)-decodable rateless matrix \(M_{n=8,d=2}\), given as follows:
\[
M_{8,2} =
\begin{bmatrix}
1 & 1 & 1 & 1 & 1 & 1 & 1 & 1 \\
0 & 1 & 0 & 1 & 0 & 1 & 0 & 1 \\
0 & 0 & 1 & 1 & 0 & 0 & 1 & 1 \\
0 & 0 & 0 & 0 & 1 & 1 & 1 & 1 \\
\end{bmatrix}
\]
To observe that $M_{8,2}$ is indeed a \(2\)-decodable rateless matrix, note that the matrix $M_{8,2}$ is \(2\)-decodable since any set of two columns has a row of weight one. Hence, \(m_2 = 4\). Furthermore, \(m_1 = 1\) since the submatrix $M_1$ of $M_{8,2}$ which consists of its first row, i.e. 
$
M_{1} =
\begin{bmatrix}
1 & 1 & 1 & 1 & 1 & 1 & 1 & 1 \\
\end{bmatrix}
$ is \(1\)-decodable because any set containing one column has a row of weight one. In summary, the matrix $M_{8,2}$ is a \(2\)-decodable rateless matrix and its decodability profile is $(m_1,m_2)=(1,4)$.

Next, we present several CertainSync constructions, where for each construction, we present a $d$-decodable rateless matrix that is used to map elements to IBLT cells differently. For each construction, an example of a $d$-decodable rateless matrix is provided in Appendix \ref{appx:additional_examples}.

\subsection{Construction \RNum{1}: EGH Rateless Matrix}
\label{egh_construction}
In this section, we show a construction of a $d$-decodable rateless matrix using the EGH method \cite{EGH}. 
\begin{definition}\label{def:EGH Matrix}[The EGH Matrix]
Let \( n \in \mathbb{N} \) and \( i \in [d] \) be given, where \( n \) is the number of columns and \( i \) is the decodability parameter. The $i$-decodable EGH matrix, denoted by \( M^{\RNum{1}}_{n,i} \in \{0,1\}^{m_i \times n} \), is defined as follows:
\begin{enumerate}
    \item Let $k_i$ be the smallest integer s.t.  $\Pi_{k_i} \geq n^i$, where $\Pi_{k_i}$ is the product of the first $k_i$ prime numbers.
    \item Let \( m_i = \sum_{j=1}^{k_i} p_j \).
    \item For each \( j \in [k_i] \), define the submatrix \( M_j \in \{0,1\}^{p_j \times n} \) as:
    \[
    M_j[x,y] = 
    \begin{cases} 
        1 & \text{if } y+1 \equiv x \pmod{p_j} \\
        0 & \text{otherwise}
    \end{cases} 
    \]
    for \( x \in [p_j]_0 \) and \( y \in [n]_0 \).
    \item The EGH matrix \( M^{\RNum{1}}_{n,i} \) is the vertical concatenation of the $k_i$ submatrices $M_1, M_2, \ldots, M_{k_i}$. 
\end{enumerate}
The resulting matrix \( M^{\RNum{1}}_{n,i} \) is of size \( m_i \times n \), where \( m_i = \sum_{j=1}^{k_i} p_j \). The value $m_i$ of the number of rows is also denoted by $m(n, i)$.
\end{definition}

In order to show that the EGH matrix $M^{\RNum{1}}_{n,d}$ is a $(d{+}1)$-decodable rateless matrix, we use the following theorem
 whose proof is deferred to Appendix \ref{appx:constructions_proofs}.  

\begin{theorem}
For $n$ and $d$, the EGH matrix $M^{\RNum{1}}_{n,d}$ is a $(d{+}1)$-decodable rateless matrix. For $2\leq i\leq d+1$, the decodability profile is
$ m_i = \sum_{j=1}^{k_i} p_j,$
where $k_i$ is the smallest integer such that $\Pi_{k_i}\geq n^{i}$. Furthermore, $m_1=m_2$.
\end{theorem}

\subsection{Construction
\RNum{2}: OLS Rateless Matrix}
\label{ols_construction}
In this section, we present a construction of a $\ceil{\sqrt{n}}$-decodable rateless matrix using the OLS method based on \cite{OLS} using a combination of mutually orthogonal Latin squares and the aforementioned special matrix $R_n$, which we interchangeably refer to as $L_0$. We present the formal definition. 

\begin{definition}\label{def:OLS Matrix}[The OLS Matrix]
Let \( n \in \mathbb{N} \) and \( i \in [\ceil{\sqrt{n}}] \) be given, where \( n \) is the number of columns and \( i \) is the decodability parameter. The $i$-decodable OLS matrix, denoted by  \( M^{\RNum{2}}_{n,i} \in \{0,1\}^{m_i \times n} \), is defined as follows:
\begin{enumerate}
    \item Let \( s = \lceil \sqrt{n} \rceil \) be a prime power and \( m_i = i \cdot s \).
    \item For each \( j \in [i]_0 \), define the submatrix \( M_j \in \{0,1\}^{s \times n} \) as:
    \begin{enumerate}
        \item For $j > 0$, use Latin square $j$ of order \( s \), or  the special matrix $R_s$ for $j = 0$, such that $ R_s \cup \{L_j \mid 1 \leq j < i\}$ is a mutually orthogonal set.
        \item For each \( k \in [n]_0 \):
        \begin{enumerate}
            \item \( x = \left\lfloor \frac{k}{s} \right\rfloor \) and \( y = k \bmod s \).
            \item 
            \[
            M_j[x,k] = \begin{cases} 
                1 & \text{if } x = L_j[x, y]  \\
                0 & \text{otherwise}
            \end{cases}
            \]
        \end{enumerate}
    \end{enumerate}
    \item The OLS matrix \( M^{\RNum{2}}_{n,i} \) is the vertical
    concatenation of the $i$ submatrices $M_0, M_1, \ldots M_{i-1}$. 
\end{enumerate}
The resulting matrix \( M^{\RNum{2}}_{n,i} \) is of size \( m_i \times n \), where \( m_i = i \cdot s \).
\end{definition}

In order to show that the OLS matrix $M^{\RNum{2}}_{n,d}$ is a $\ceil{\sqrt{n}}$-decodable rateless matrix, we use the following theorem  whose proof is deferred to Appendix \ref{appx:constructions_proofs}.  

\begin{theorem}
\label{tho:ols}
For $n$ where $\ceil{\sqrt{n}}$ is a prime power, and $d=\ceil{\sqrt{n}}$, the OLS matrix $M^{\RNum{2}}_{n,d}$ is a $\ceil{\sqrt{n}}$-decodable rateless matrix. For $1\leq i\leq \ceil{\sqrt{n}}$, the decodability profile is
$m_i =  i \cdot \ceil{\sqrt{n}}$.
\end{theorem}

\subsection{Construction \RNum{3}: Extended Hamming Rateless Matrix}
\label{extended_hamming_construction}
In this section, we present a construction of a 3-decodable rateless matrix based on Extended Hamming code construction from \cite{EXTENDED_HAMMING_CODE}.  
For any positive integer $n$ (not necessarily a power of 2), we let
\(m = \lceil \log_2(n) \rceil\). 
The matrix $H_n^{'}$ consists of the first $n$ columns of all the $2^m$ binary column vectors of the matrix $H_{2^m}^{'}$. Thus, $H_n^{'}$ has dimensions $m \times n$, and $H_n$ is also similarly extended to have dimensions $(m+1) \times n$ by adding the all ones row to the matrix $H_n^{'}$.
This ensures that for any $n$, we use the parity check matrix of the extended Hamming code for the smallest power of 2 greater than or equal to $n$, and then take only the first $n$ columns of that matrix.

\begin{definition}\label{def:Extended_Hamming_Matrix}[The Extended Hamming Matrix]
Let \( n \in \mathbb{N} \) be given, where \( n \geq 8 \) is the number of columns. The $3$-decodable Extended Hamming matrix, denoted by \( M^{\RNum{3}}_{n,3} \), is defined as follows:
\[
M^{\RNum{3}}_{n,3} = \begin{bmatrix}
H_n \\
\overline{H_n^{'}}
\end{bmatrix},
\]
where the matrix $\overline{H_n^{'}}$ is the binary complement of the matrix $H_n^{'}$. The resulting matrix \( M^{\RNum{3}}_{n,3} \) is of size \( m_3 \times n \), where \( m_3 = 2 \lceil \log_2 n \rceil + 1 \). 
\end{definition}

In order to show that the Extended Hamming matrix $M^{\RNum{3}}_{n,d}$ is a $3$-decodable rateless matrix, we use the following theorem  whose proof is deferred to Appendix \ref{appx:constructions_proofs}.

\begin{theorem}
\label{tho:ed}
For $n\geq 8$ and $d = 3$, the Extended Hamming matrix $M^{\RNum{3}}_{n,d}$ is a 3-decodable rateless matrix, where the decodability profile is $(m_1,m_2,m_3)=(1,\lceil \log_2 n \rceil+1,2\lceil \log_2 n \rceil+1)$. 
\end{theorem}

\section{Experimental Evaluation}
\label{sec:experimental_evaluation}
This section examines our proposed CertainSync constructions compared to baseline schemes for set reconciliation and demonstrates the construction's comparable performance to the state-of-the-art scheme, Rateless IBLT \cite{RATELESS_IBLT}, under different configurations for each construction (universe size, symmetric difference size, and communication overhead).

\textbf{Schemes Comparison.} We compare our CertainSync constructions (EGH, OLS, and Extended Hamming) which do not require any parametrization or estimators, to other set reconciliation schemes. We evaluate our constructions alongside state-of-the-art Rateless IBLT~\cite{RATELESS_IBLT} as a rateless scheme, and Difference Digest~\cite{EFFICIENT_SET_RECONCILIATION} and Graphene~\cite{GRAPHENE} as non-rateless schemes. In particular, Difference Digest incurs additional communication overhead due to the transmission of a Strata Estimator, which is required to estimate the size of the symmetric difference before reconciliation.
\begin{figure}[b!]
\centering
\begin{minipage}[t]{0.34\textwidth}
\centering
\includegraphics[width=\textwidth]{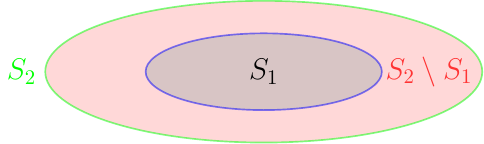}
\caption*{(a) Simplified case where $S_2$ is a superset of $S_1$ ($S_1 \subseteq S_2$).}
\end{minipage}
\hfill 
\begin{minipage}[t]{0.40\textwidth}
\centering
\includegraphics[width=\textwidth]{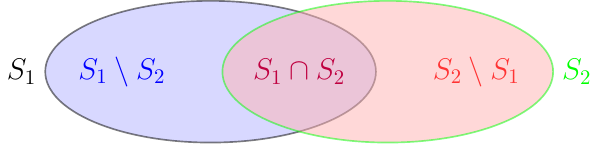}
\caption*{(b) General case with arbitrary set overlap between $S_1$ and $S_2$.}
\end{minipage}
\caption{Set scenarios between two participants.}
\label{fig:venn_sets_ relationship}
\end{figure}

\textbf{Setup.} In the experiments, we refer to the common simplified scenario that one set is a superset of another set as a typical case, as observed in prior work~\cite{GRAPHENE}, which simplifies the computation of the symmetric difference $\Delta$ by merely removing elements of one set from the other. Later, in Section \ref{sec:universe_reduce_sync_framework}, which addresses the blockchain applicability of CertainSync, this assumption is not in use anymore. Fig. \ref{fig:venn_sets_ relationship} illustrates two set scenarios. In our experimental setup, each IBLT cell comprises three 8-byte fields (counter, xorSum and checkSum), totaling 24 bytes or 192 bits (24 bytes $\times$ 8) per cell across all evaluated schemes to ensure a fair comparison. To assess metrics of set reconciliation schemes such as accuracy, communication efficiency, and scalability, we developed an experimental design wherein $10$ independent trials were conducted for each scheme. Specifically, for each trial, the set of the receiver of IBLT cells was defined as the complete universe of $n$ elements. In contrast, the set of the sender of IBLT cells was constructed by removing $|\Delta|$ (symmetric difference size) unique, randomly selected elements from the receiver set. The experimental results were analyzed by averaging the metrics across the trials, thereby mitigating potential bias in results due to random removal of elements from the sender set. The additional communication overhead of sending a subset of \( \Delta \) elements from the receiver to the transmitter at the end is neglected, as it does not involve the transmission of IBLT cells and is treated as a constant additive factor across all schemes.

\begin{figure}[h!]
\centering
\begin{minipage}[t]{0.35\textwidth}
\centering
\includegraphics[width=\textwidth]{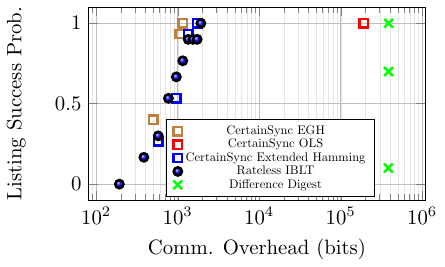}
\caption*{(a) $m(n=10^{6}, |\Delta|=3)$}
\end{minipage}
\hfill 
\begin{minipage}[t]{0.31\textwidth}
\centering
\includegraphics[width=\textwidth]{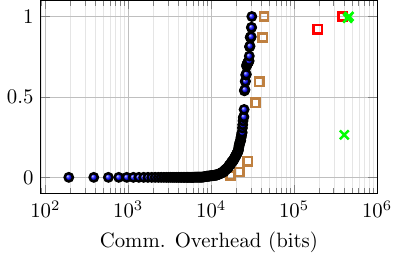}
\caption*{(b) $m(n=10^{6}, |\Delta|=100)$}
\end{minipage}
\hfill 
\begin{minipage}[t]{0.30\textwidth}
\centering
\includegraphics[width=\textwidth]{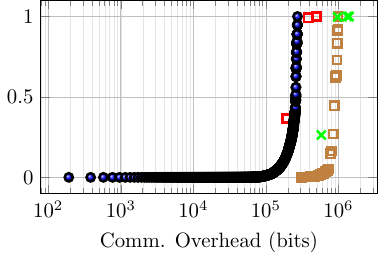}
\caption*{(c) $m(n=10^{6}, |\Delta|=1000)$}
\end{minipage}
\caption{The trade-off between listing success probability for various schemes as a function of communication overhead (in bits). The universe size is $n =10^6$ and $|\Delta|$ refers to the symmetric difference size.} 
\label{fig:success_rate_vs_total_bits_set_inside_set}
\end{figure}

\textbf{Findings.}
Our experimental results demonstrate that our CertainSync constructions, which are parameterless with a guarantee for successful set reconciliation, achieve comparable performance to state-of-the-art Rateless IBLT scheme across the different configurations, with each construction exhibiting optimal performance for specific symmetric difference size ranges. Specifically, CertainSync EGH and Extended Hamming excel with small to medium symmetric difference size ranges, requiring minimal communication overhead similar to Rateless IBLT, while CertainSync OLS becomes more efficient as the symmetric difference size approaches \( \lceil \sqrt{n} \rceil \). The experiments reveal clear trade-offs between decoding accuracy, communication efficiency, and scalability across all schemes, with Rateless IBLT consistently demonstrating superior overall performance but CertainSync constructions offering competitive alternatives for specific use cases as parameterless solutions that eliminate the need for estimators.

\subsection{Decoding Accuracy}
Fig. \ref{fig:success_rate_vs_total_bits_set_inside_set} presents the listing success probability of various schemes as a function of the communication overhead for different symmetric difference sizes in a linear-log plot. For small symmetric difference sizes, such as \( |\Delta| = 3 \), CertainSync (EGH and Extended Hamming) and Rateless IBLT demonstrate the lowest communication overhead to reach a success probability of 1. In contrast, CertainSync OLS and Difference Digest exhibit the poorest efficiency with almost double the communication overhead of previous schemes to reach a probability of 1. 
As the symmetric difference size increases to medium levels (\( |\Delta| = 100 \)), Rateless IBLT maintains superior performance, closely followed by CertainSync EGH. However, CertainSync OLS and Difference Digest continue to underperform, requiring approximately ten times the communication overhead compared to the former schemes to achieve a probability of 1.
In the large symmetric difference size regime (\( |\Delta| = 1000 \)), Rateless IBLT remains the most efficient, achieving a success probability converging to 1 as the number of IBLT cells approaches \( 1.35 |\Delta| \). CertainSync OLS shows comparable performance, particularly at lower success probabilities. CertainSync EGH becomes notably inefficient, demanding approximately four times the communication overhead of Rateless IBLT to reach a success probability of 1. Difference Digest maintains the worst efficiency compared to other schemes and reaches a probability of 1 with relatively the same communication overhead as CertainSync EGH. 
The experimental results reveal a trade-off between symmetric difference size and communication overhead for CertainSync constructions. Each CertainSync construction exhibits an optimal range of symmetric difference sizes, achieving performance comparable to the Rateless IBLT scheme, which demonstrated the best overall performance. On one hand, CertainSync OLS requires,  compared to CertainSync EGH, excessive communication overhead for small to medium symmetric difference sizes (\( |\Delta| \ll \lceil \sqrt{n} \rceil = 1000 \)), due to its dependence on \( \lceil \sqrt{n} \rceil \). However, as the symmetric difference size increases, as illustrated for \( |\Delta| = 1000 \), CertainSync OLS achieves a success probability converging to 1 with significantly lower communication overhead compared to CertainSync EGH, which becomes increasingly inefficient with increasing symmetric difference size.  

\subsection{Communication Efficiency}
\begin{figure}[t]
\centering
\begin{minipage}[t]{0.410\textwidth}
\centering
\includegraphics[width=\textwidth]{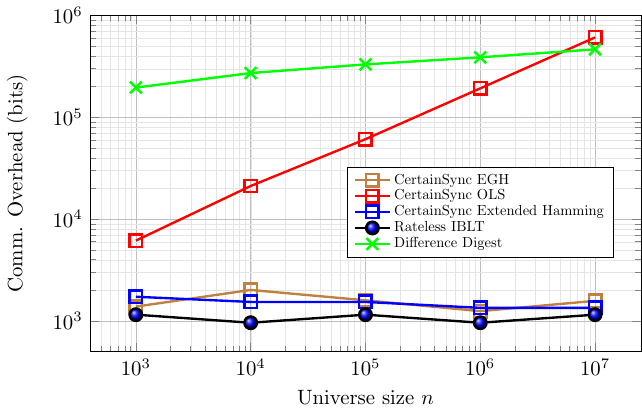}
\caption*{(a) $m(n, |\Delta|=3)$}
\end{minipage}
\hfill 
\begin{minipage}[t]{0.380\textwidth}
\centering
\includegraphics[width=\textwidth]{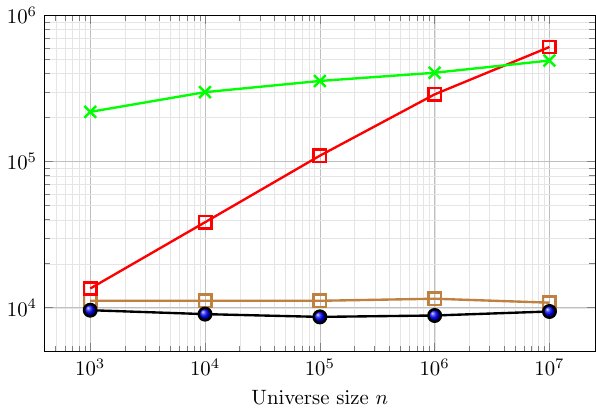}
\caption*{(b) $m(n, |\Delta|=30)$}
\end{minipage}
\caption{Trade-off between communication overhead  for various schemes as a function of universe size \( n \).}
\label{fig:total_bits_vs_universe_size_set_inside_set}
\end{figure}

\begin{figure}[t!]
    \centering
    \begin{minipage}[t]{0.460\textwidth}
        \centering
        \includegraphics[width=\textwidth]{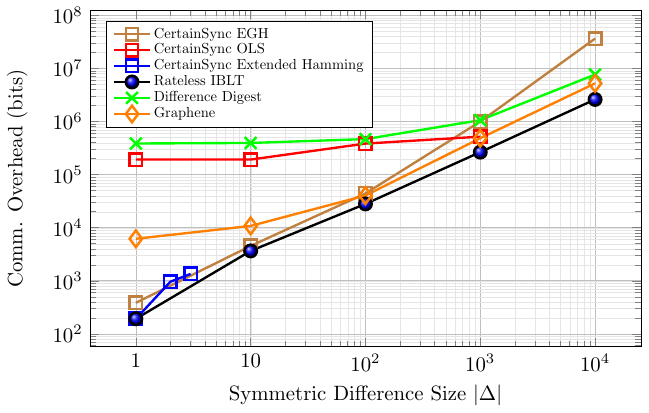}
        
        \caption*{(a) Communication Overhead (in bits)}
    \end{minipage}
    \hfill
    \begin{minipage}[t]{0.460\textwidth}
        \centering
        \includegraphics[width=\textwidth]{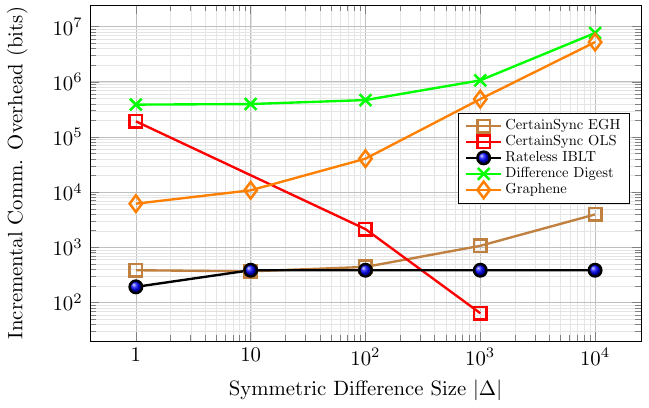}
        
        \caption*{(b) Incremental Communication Overhead (in bits)}
    \end{minipage}
    \caption{Communication overhead and incremental communication overhead (in bits) for various schemes as a function of symmetric difference size \( |\Delta| \). The universe size is $n =10^6$.}
    \label{fig:communication_overhead}
\end{figure}

Fig.~\ref{fig:total_bits_vs_universe_size_set_inside_set} shows the communication overhead of various schemes as a function of the universe size $n$ in a log-log plot.
We focus on a small symmetric difference size (\( |\Delta| = 3 \)), and also evaluate  larger symmetric difference  size
($|\Delta| = 30$) that approaches the theoretical upper bound $|\Delta_{\text{max}}| = \ceil{\sqrt{10^3}} = 32$ imposed by CertainSync OLS for minimal universe size $n = 10^3$ experimented with.
For $|\Delta| = 3$, Rateless IBLT and CertainSync (EGH and Extended Hamming)  exhibit optimal performance with approximately constant growth regardless of the universe size, and require the least communication overhead. 
For Rateless IBLT, the communication overhead,  theoretically based on \cite{RATELESS_IBLT}, lies between \( 1.35|\Delta| \) and \( 1.72|\Delta| \) on average, confirming that the complexity with respect to the universe size is \( O(1) \). In contrast, the theoretical communication overhead complexity for CertainSync Extended Hamming is \( O(\log n) \), while for CertainSync EGH it follows \( O\left( \log^2 n\right) \). Simulation results demonstrate that both constructions achieve practical performance consistent with \( O(1) \) complexity, reflecting their dependency on the symmetric difference size (\( |\Delta| \)) more than the universe size.
CertainSync OLS demonstrates a worse performance in comparison to the previous schemes, as it is more susceptible to changes in universe size. Specifically,  CertainSync OLS requires communication overhead proportional to \( \lceil \sqrt{n} \rceil \), which increases significantly with larger universe sizes. The Difference Digest demonstrates the worst performance in comparison to the other schemes, with the highest communication overhead requirements, even for a small universe size, due to sending an estimator before reconciliation. 
Fig. \ref{fig:communication_overhead}(a) presents the communication overhead of various schemes as a function of the symmetric difference size $|\Delta|$ in a log-log plot. 
For low symmetric difference sizes $(|\Delta| < 10)$, Rateless IBLT demonstrates the best performance, requiring the least communication overhead, closely followed by CertainSync constructions of EGH and Extended Hamming.  Graphene shows moderate efficiency, while  CertainSync OLS, with its initial transmission of $\ceil{\sqrt{n}}$ cells, and Difference Digest with the transmission of an estimator, perform poorly in this range by having the highest overhead.
For medium symmetric difference sizes $(10 < |\Delta| < 1000)$, Rateless IBLT maintains its superior performance with the lowest communication overhead; Graphene achieves comparable overhead, while CertainSync EGH demonstrates moderate performance. As CertainSync OLS approaches its maximum symmetric difference size $|\Delta|_{max} = \ceil{\sqrt{10^6}} = 10^3$,  its performance is better than CertainSync EGH. Difference Digest still with the worst performance due to the initial transmission of an estimator.
For high symmetric difference sizes $(|\Delta| > 1000)$, Rateless IBLT remains the most efficient scheme, requiring the least communication overhead. Graphene performs similarly well but is slightly less efficient than Rateless IBLT.
Difference Digest demonstrates moderate performance, and the overhead of transmission of IBLT cells after the estimation phase is comparable to the communication overhead of estimation, thus a rise in overhead as symmetric difference size increases. 
CertainSync EGH demonstrates the poorest scalability, requiring significantly more communication overhead than all other schemes. Theoretical communication costs almost align with these results. Rateless IBLT achieves $O(|\Delta|)$ complexity, with simulations confirming overhead between $1.35|\Delta|$ and $1.72|\Delta|$ on average, converging to $1.35|\Delta|$ for larger $|\Delta|$. CertainSync EGH, with theoretical complexity $O(|\Delta|^2)$, scales poorly for large $|\Delta|$. CertainSync Extended Hamming exhibits $O(|\Delta|)$ complexity, which makes it effective for small symmetric difference sizes, while CertainSync OLS demonstrates parts of $O(1)$ and $O(|\Delta|)$, which implies superior scaling in parts of $O(1)$ due to a stronger dependence on universe size $n$. It performs well for medium values of $|\Delta|$. Difference Digest achieves $O(|\Delta|)$ complexity through Strata Estimator based estimation of $|\Delta|$, and Graphene, despite high initial overhead for small $|\Delta|$  as much as 20\% higher than its true minimum cost (see Eq. $3$ in \cite{GRAPHENE}), converges to $O(|\Delta|)$ for larger symmetric difference sizes. 

\subsection{Scalability}
In this context, scalability is measured by the rate at which incremental communication overhead (additional bits) increases concerning the size of the symmetric difference while maintaining a constant universe size. Schemes requiring fewer additional bits as the symmetric difference size grows are considered more scalable. A steeper slope indicates poorer scalability, as the communication overhead increases rapidly. Conversely, a flatter slope indicates better scalability, as the scheme requires fewer additional bits for larger symmetric difference sizes.
Fig. \ref{fig:communication_overhead}(b) shows the incremental communication overhead of various schemes as a function of symmetric difference size $|\Delta|$ in a log-log plot. In this simulation, the distinction between rateless and non-rateless schemes becomes evident. The incremental overhead of non-rateless schemes, such as Difference Digest and Graphene, is the same as the total communication overhead (see Fig. \ref{fig:communication_overhead}(a)) due to their inability to dynamically expand, constrained by either fixed IBLT size based on parametrization (Graphene), or estimation-based allocation (Difference Digest). Consequently, their overall scalability is significantly inferior compared to rateless schemes.
Rateless IBLT demonstrates superior overall scalability, exhibiting constant growth in incremental communication overhead as the symmetric difference size increases above $|\Delta| = 10$. CertainSync EGH achieves comparable overhead to Rateless IBLT up to medium symmetric difference size of $|\Delta| = 100$, and beyond it exhibits linear growth, becoming less scalable with complexity $O(|\Delta|)$. CertainSync OLS presents relatively high incremental communication overhead for small symmetric difference sizes due to its dependence on universe size $n$, performing worse than even the non-rateless Graphene scheme. However, it demonstrates improved scalability compared to non-rateless schemes at medium symmetric difference sizes, and notably, as it approaches its maximum symmetric difference size $|\Delta|_{\text{max}} = \ceil{\sqrt{10^6}} = 10^3$, its incremental overhead surpasses that of Rateless IBLT.

\section{C\lowercase{ertain}S\lowercase{ync} for Blockchain Synchronization}

\label{sec:universe_reduce_sync_framework}

\subsection{Setup}

We utilize an architecture comprising two Ethereum blockchain nodes, each consisting of an execution client and a consensus client.
The execution client used is Geth, which facilitates the processing of transactions and the execution of smart contracts on the Ethereum blockchain. The consensus client employed is Prysm, which implements Ethereum's Proof of Stake (PoS) consensus mechanism. Each node operates independently but participates in the same network. By utilizing the Geth \textit{admin.peers} API, we can confirm that these nodes are not peers of one another. This separation is advantageous, as it might lead to more distinct transaction pools for each node as we mainly focus on synchronization of transactions rather than blocks. For this experiment, we utilize the Sepolia testnet, which is one of the Ethereum test networks. Sepolia provides a sandbox environment that allows testing applications and smart contracts without incurring real costs or risks associated with the main Ethereum network (Mainnet). 
\begin{figure}[h!]
    \centering
    \begin{minipage}[t]{0.430\textwidth}
        \centering
        \includegraphics[width=\textwidth]{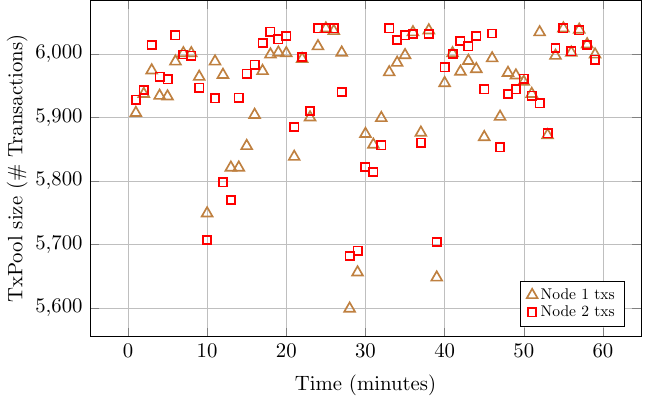}
        
        \caption*{(a) The size of the Ethereum transaction pools (TxPools), combining both queued and pending transactions over time (in minutes).}

    \end{minipage}
    \hfill
    \begin{minipage}[t]{0.430\textwidth}
        \centering
        \includegraphics[width=\textwidth]{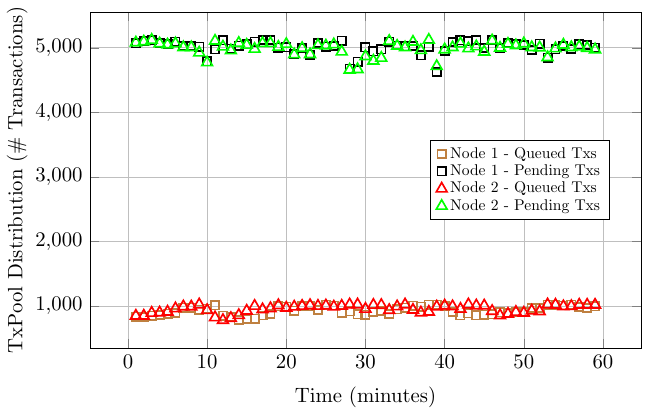}
        
        \caption*{(b) The count of queued and pending transactions in transaction pools (TxPools) over time (in minutes).}
    \end{minipage}
    \caption{Characteristics of Ethereum transaction pools (TxPools).}
    \label{fig:nodes_txs_characteristics}
\end{figure}
In blockchain, the transaction pool (TxPool), analogous to the Mempool in Bitcoin, is a critical component that holds all transactions that have been submitted but not yet included in a block. 
We collected and analyzed the real content of the TxPools at two Ethereum nodes on November 27, 2024, for an hour (which ended around noon EST). 
The TxPool is divided into two types: queued transactions, which wait for processing and inclusion in a block, prioritized by gas price for miners' selection, and pending transactions, which have been selected by a miner and are in the process of being included in a block but are not yet confirmed or added to the blockchain.
For the purpose of synchronizing transaction pools, we place less emphasis on transaction types, but rather on the total number of them at any given time as shown in Fig. \ref{fig:nodes_txs_characteristics}(a).  We stick to the default maximum values as specified in the Geth client version 1.14.11-stable-f3c696fa, where the maximum pending transactions (controlled by \texttt{txpool.globalslots}) is limited to 1024 transactions, the maximum queued transactions (controlled by \texttt{txpool.globalqueue}) is limited to 5120 transactions, and queued transactions are removed after 3 hours (controlled by \texttt{txpool.lifetime}). After giving each node enough time to be fully synchronized to the chain, 
from Fig. \ref{fig:nodes_txs_characteristics}(b), we can observe that the number of queued transactions is indeed close to 1024 with declines due to the discarding of queued transactions, or due to becoming a pending transaction, while pending transaction count is around 5000 with declines due to appending transactions to a block, and increases due to queued transactions    becoming pending. By utilizing this setup, we intend to explore transaction pool synchronization between the two nodes.  

\subsection{The UniverseReduceSync Framework}

\subsubsection{Motivation.}

Each transaction in a blockchain network includes a hash field, which is generated using a cryptographic hash function like SHA256. This function processes the transaction details, such as the transaction value and the sender's and receiver's addresses, to produce a unique identifier. In the Ethereum network, a transaction includes a hash field of 256 bits. At first glance, this implies a universe size of $n = 2^{256}$ and substantial inefficiency in the basic CertainSync framework, with communication overhead affected by the universe size (Table \ref{table_constructions}). For example, in CertainSync OLS construction, communication overhead is proportional to $\ceil{\sqrt{n=2^{256}}} = 2^{128}$. 
Therefore, we extend our CertainSync framework to incorporate universe size reduction, resulting in an extended framework for large-scale universe size scenarios such as blockchain networks, which we refer to as \emph{UniverseReduceSync}. By reducing the universe size to a scale proportional to the actual size of the sets involved in synchronization, such as the
size of the transaction pools in blockchain networks as demonstrated by using the Ethereum blockchain as a case study, we significantly reduce the communication overhead compared to the basic CertainSync framework.

\begin{figure}[h!]
    \centering
    \begin{minipage}[t]{0.38\textwidth}
        \centering
        \includegraphics[width=\textwidth]{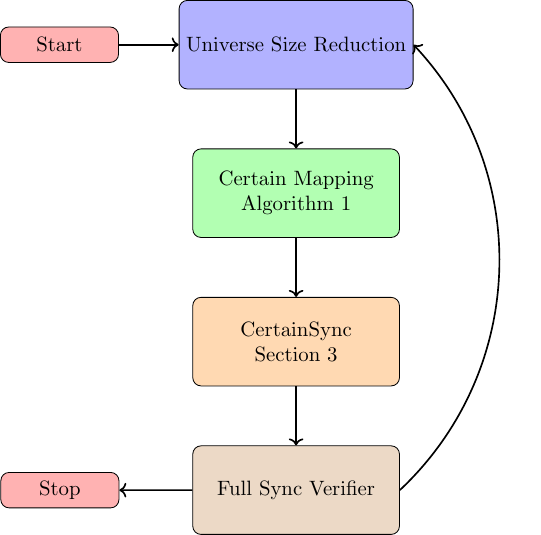}
        
        \caption*{(a) Block diagram illustrating the architecture and workflow of the UniverseReduceSync.}

    \end{minipage}
    \hfill
    \begin{minipage}[t]{0.564\textwidth}
        \centering
        \includegraphics[width=\textwidth]{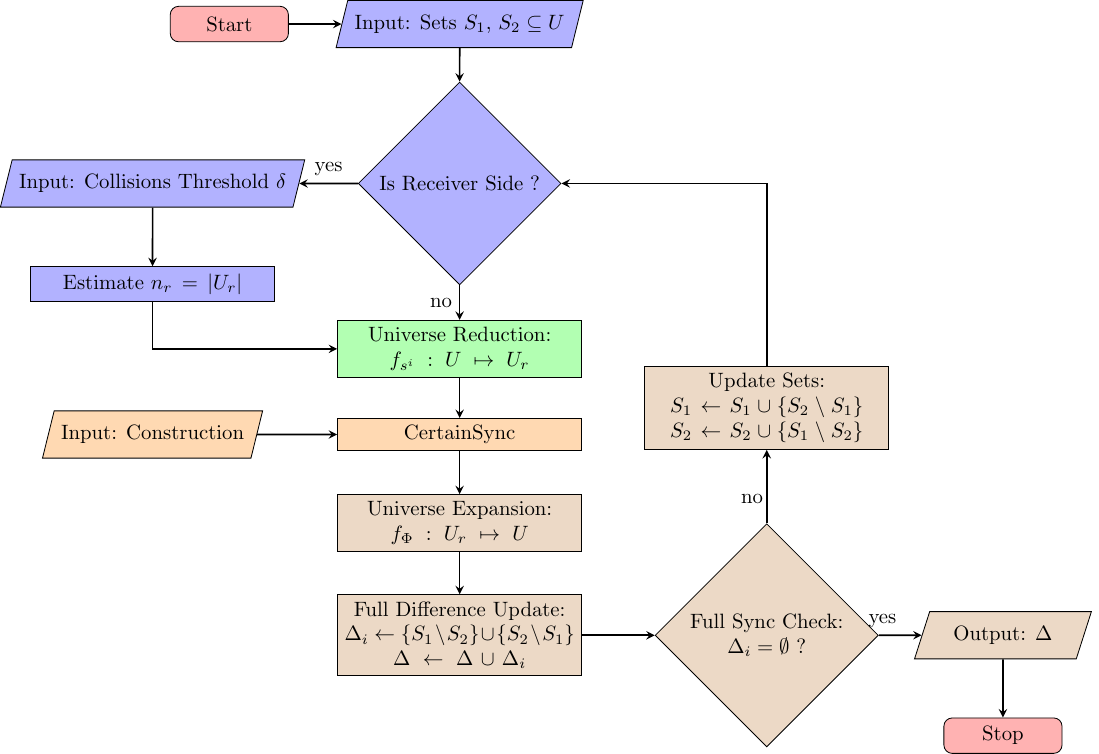}
        
        \caption*{(b) The detailed flowchart for UniverseReduceSync. The colors match the components in the block diagram.}
    \end{minipage}
    \caption{UniverseReduceSync illustration via block diagram and flowchart.}
    \label{fig:universe_reduce_sync_flow}
\end{figure}

\subsubsection{Architecture} 

In UniverseReduceSync, as described in Fig. \ref{fig:universe_reduce_sync_flow}, there is a cyclic process designed to facilitate synchronization in multiple rounds, where in each round CertainSync is used. The key components of the framework are the following:

\textbf{UniverseSizeReduction.}
This component estimates the new reduced universe size ($n_r$) based on the total elements from the original universe size $n$ of sets $S_1$ and $S_2$. Its estimation aims to minimize the number of collisions in the reduced universe below a threshold denoted as $\delta$. 

\textbf{Certain Mapping.}
In this component, each element $e$ in a set undergoes a mapping using a predefined hash function $\mathcal{H}$ parameterized by a hash salt \( s^{i} \), where \( i \) denotes the round number. The hash salt \(s^i\) is generated using an agreed pseudorandom number generator (PRNG), ensuring consistent and reproducible values for \(s^i\) among all participants. Specifically, for each element $e$, we compute a reduced element value \( e_r = \mathcal{H}(e, s^{i}) \). Additionally, it constructs a reverse mapping $\Phi$ from the reduced universe back to the original universe, ensuring that each element in the reduced universe can be mapped back to its corresponding original element or elements in case of duplicates.

\textbf{CertainSync.}
The CertainSync component performs the synchronization process within the reduced universe, leveraging the reduced elements $S_r$ from the previous step on each side.

\textbf{Fully Sync Verification.}
The Fully Sync verification component determines whether the synchronization is fully done. If so, the process yields the symmetric difference $\Delta$. However, if the synchronization is not fully done, the process returns to the Universe Size Reduction stage with less data to synchronize than the previous round. 

\subsubsection{Trade-offs.} Table \ref{tab:sync_comparison} illustrates the key trade-offs between CertainSync and UniverseReduceSync frameworks. For CertainSync, EGH is the sole applicable construction, as it avoids the limitations of Extended Hamming with a small maximum symmetric difference size ($|\Delta_{max}|=3$), and is significantly less dependent on the universe size $n$  with polylogarithmic communication overhead complexity $O(\log^2 n)$, compared to OLS proportional to $\lceil \sqrt{n} \rceil$. In UniverseReduceSync, while Extended Hamming remains irrelevant due to its small maximum symmetric difference size, both EGH and OLS constructions prove applicable. Moreover,
UniverseReduceSync achieves lower communication costs through reduced elements but introduces potential collisions and requires parameter tuning and estimation of the reduced universe size. In contrast, CertainSync offers simpler, collision-free operation with guaranteed single-round synchronization at the expense of higher communication overhead due to the use of elements in the original large universe. 

\begin{table}[h]
\centering
\caption{Framework Characteristics: CertainSync vs. UniverseReduceSync. \textcolor{green}{Green} indicates a preferable characteristic, while \textcolor{red}{red} highlights a less favorable one.}
\resizebox{\textwidth}{!}{\begin{tabular}{|c|c|c|}
\hline
\backslashbox{Property}{Framework} & \makecell{CertainSync \\ (Section \ref{sec:certain_sync_framework})} & \makecell{UniverseReduceSync \\ (Section \ref{sec:universe_reduce_sync_framework})} \\
\hline
Constructions & {\color{red}EGH} & {\color{green}EGH, OLS} \\
\hline
Universe Size & {\color{red}Original} & {\color{green}Reduced} \\
\hline
Communication Complexity & {\color{red}Higher} & {\color{green}Lower} \\
\hline
Parametrization \& Estimation & {\color{green}None} & {\color{red}Includes} \\
\hline
Collisions & {\color{green}No collisions} & {\color{red}Possible within and between participants} \\
\hline
Synchronization Rounds & {\color{green}Exactly 1 round} & {\color{red}$\geq 1$ rounds (collision-dependent)} \\
\hline
Memory Overhead & 
{\color{green} Original elements only} &
{\color{red} Original and reduced elements} \\
\hline
\end{tabular}}
\label{tab:sync_comparison}
\end{table}

\subsection{Design of UniverseReduceSync}

\textbf{UniverseSizeReduction.} The reduced universe size $n_r$ must be at least $2^{\ceil{\log_2(m)}}$ where $m = |S_1| + |S_2|$ serves as an upper bound on $|S_1 \cup S_2|$ due to potential overlaps of same elements. This bound ensures: \emph{(i)} avoiding guaranteed collisions by the pigeonhole principle when $n_r < m$, \emph{(ii)} optimal bit representation since any $n_r < 2^{\ceil{\log_2(m)}}$ would not fully utilize the minimum bits needed for $m$ distinct values. While $2^{\ceil{\log_2(m)}}$ provides sufficient capacity for unique representations, the actual number of collisions depends on the statistical properties of the chosen hash used to map to the reduced universe, the reduced universe size $n_r$ and the hash salt per round $s_i$. 

\begin{theorem}[Hash Collisions Expectation]
Let $\mathcal{H}$ be a hash function that maps elements to a universe of size $n_r$, and let $m$ be the number of elements. Assuming a uniform hash distribution, the expected number of element collisions is $E[\text{Collisions}] = \frac{m(m-1)}{2n_r}$.
\end{theorem} 
\begin{proof} 
The probability of a specific pair's collision is $\frac{1}{n_r}$, with $\binom{m}{2} = \frac{m(m-1)}{2}$ total possible element pairs. Consequently, the expected number of element collisions is $\binom{m}{2} \cdot \frac{1}{n_r} = \frac{m(m-1)}{2n_r}$.
\end{proof}

To determine the minimal value of \( n_r \), we start with the constraint that the expected number of collisions, given by \( E[\text{collisions}] = \frac{m(m-1)}{2 \cdot n_r} \), must not exceed \( \delta \). Through algebraic manipulation of this inequality, we find that \( n_r \geq \frac{m(m-1)}{2\delta} \). Since \( n_r \) must be a positive integer, we take the ceiling function of this expression, which yields \( n_r = \left\lceil \frac{m(m-1)}{2\delta}\right\rceil \) as the minimal value. 

\textbf{CertainMapping.} Given two sets $S_1$ and $S_2$, if elements $e_1 \in S_1$ and $e_2 \in S_2$ map to the same reduced element $e_r$ under CertainMapping, then either $e_1 = e_2$ or a collision occurred.  

\begin{algorithm}[h!]
\caption{CertainMapping}
\DontPrintSemicolon
\SetKwInOut{Input}{Input}
\SetKwInOut{Output}{Output}
\Input{Elements $S \subseteq U$, Hash Salt $s^{i}$, Reduced Universe size $n_r$}
\Output{Reduced elements $S_r$, Mapping $\Phi$}
Initialize $S_r \gets \emptyset$, $\Phi \gets$ dictionary$\{ \}$\;
\For{$e \in \mathcal{S}$}{
$e_r \gets \left(\mathcal{H}(e, s^i) \bmod n_r\right) + 1$\;
$S_r \gets S_r \cup \{e_r\}$\;
$\Phi(e_r) \gets \Phi(e_r) \cup \{e\}$\;
}
\Return{$(S_r, \Phi)$}\;
\end{algorithm}


\textbf{Fully Sync Verifier.} Algorithm \ref{alg:DecodeDiff} in Appendix \ref{appx:additional_algorithms} is extended by leveraging the counter's sign in pure cells to determine the side-association of each reduced element \( e_r \in \text{IBLT}\{\Delta\} \). For the first round with an assumption for discrete uniform distribution for hash \( \mathcal{H} \), 
the symmetric difference size in this round \( |\Delta|_1 \) is bounded by \( \max(0, |\Delta| - 2\delta) \leq |\Delta|_1 \leq |\Delta| \), where the lower bound represents collisions of elements in \( \Delta \), and the upper bound represents element collisions in the intersection. Consequently, the probability of a full sync check succeeding is \( \mathcal{P}(Success) \geq \frac{\max(0, |\Delta| - 2\delta)}{|\Delta|} \). 

\subsection{Results}
\begin{figure}[h!]
\centering
\begin{minipage}[t]{0.30\textwidth}
\centering
\includegraphics[width=\textwidth]{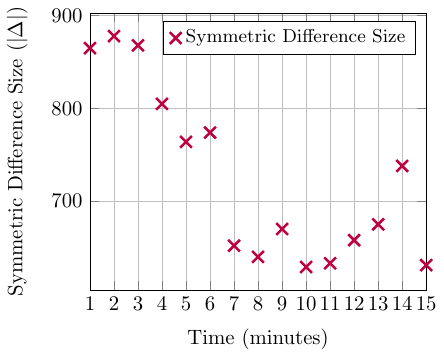}
\caption*{(a) Variation of symmetric difference size over time (in Minutes).}
\end{minipage}
\hfill 
\begin{minipage}[t]{0.30\textwidth}
\centering
\includegraphics[width=\textwidth]{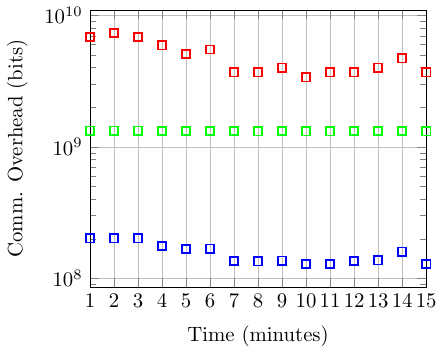}
\caption*{(b) Maximum number of collisions $\delta=1$.}
\end{minipage}
\hfill 
\begin{minipage}[t]{0.30\textwidth}
\centering
\includegraphics[width=\textwidth]{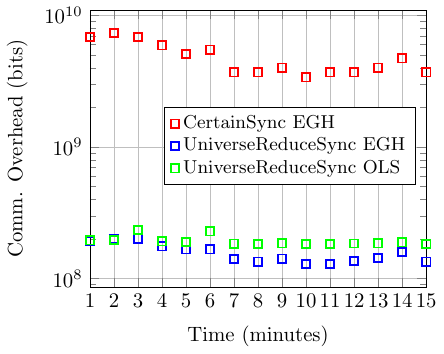}
\caption*{(c) Maximum number of collisions $\delta=100$.}
\end{minipage}
\caption{Symmetric Difference Size $|\Delta|$ and communication overhead (in bits) for various synchronization schemes as a function of maximum number of collisions ($\delta$) and time in minutes.}
\label{fig:total_bits_vs_time}
\end{figure}

To enable validation of the correctness of our results, the symmetric difference size $|\Delta|$ is required in advance, but in real time, it is unknown. We collected, as mentioned earlier, the content of TxPools at two Ethereum nodes over time at one-minute intervals, and calculated in advance their symmetric difference size as shown in Fig. \ref{fig:total_bits_vs_time}(a). The transaction unique identifier, which is the hash field, is used as the new element, in contrast to previous experiments with a positive integer. 
In our experimental setup, UniverseReduceSync constructions use an IBLT cell structure with an 8-byte counter, 32-byte xorSum, and 32-byte checkSum, totaling 72 bytes or 576 bits per cell. In contrast, CertainSync constructions utilize an IBLT cell with three 8-byte fields (counter, xorSum, checkSum), resulting in 24 bytes or 192 bits per cell. The difference in sizes is due to UniverseReduceSync utilizing original transaction hashes of 256 bits (32 bytes), whereas CertainSync utilizes reduced transaction hashes of 64 bits (8 bytes) for its xorSum and checkSum fields.

In Fig. \ref{fig:total_bits_vs_time}(b) and (c), there is a comparison of the communication overhead for three synchronization schemes: CertainSync EGH (original universe) and UniverseReduceSync EGH \& OLS (reduced universe) under varying number of collision constraints \(\delta \in \{1, 100\} \). The subplots show the communication overhead as a function of time, with CertainSync EGH consistently requiring the most overhead with the same overhead costs across \(\delta\) values, UniverseReduceSync EGH consistently requiring the least communication overhead with the same overhead costs across \(\delta\) values, and UniverseReduceSync OLS demonstrating decreased communication overhead as \(\delta\) increases, which implies a lower reduced universe size $n_r$.

\section{Conclusions and Future Work}
\looseness=-1
This paper introduced CertainSync, a novel framework for set reconciliation that guarantees success when communication overhead reaches a bound derived by the symmetric difference size and the universe size, unlike traditional schemes that offer only statistical guarantees. We proposed three rateless constructions without requiring parametrization or symmetric difference size estimation based on group testing, Latin squares, and error correction codes. We analyzed their performance and validated their effectiveness through experiments compared to other baseline schemes for set reconciliation. Additionally, we presented UniverseReduceSync, an extended framework of CertainSync for large-scale universe size reconciliation to minimize communication overhead. We evaluated it alongside the basic CertainSync framework on the Sepolia Ethereum network to compare the trade-offs between the two frameworks. A natural open question for future work refers to the development of other families of constructions for CertainSync, like combinatorial or recursive, that can be applicable to set reconciliation with certainty. We would also like to study the reconciliation of more than two sets, often known as multi-party set reconciliation, with certainty due to its practical relevance in blockchain networks. Moreover, evaluating the total reconciliation time of CertainSync compared to other baseline set reconciliation schemes and in real-world scenarios, particularly under high network churn, remains an important direction that will further validate the practical applicability.

\bibliographystyle{ACM-Reference-Format}
\bibliography{ref}

\newpage

\appendix

\section{Additional Applications of Set Reconciliation}
\label{appx:applications}

Set reconciliation is crucial in domains that require efficient data synchronization, particularly in blockchain networks. Minimizing communication overhead for synchronization between parties is essential to ensure efficiency and scalability. Some notable applications include:

\textbf{Distributed Databases.} Large-scale distributed databases like in finance often employ reconciliation to maintain consistency across replicas \cite{FINANCIAL_DATABASES_SYNC}. Also, schemes for reconciliation in distributed versioned database systems like MindPalace \cite{VERSION_RECONCILIATION}
are used for auto-mergeability, where branches may be reconciled (analogous to a branch merging in Git) without human intervention.

\textbf{Collaborative Editing.} Real-time collaborative editing systems use reconciliation techniques to synchronize user changes efficiently, ensuring consistency with minimal conflicts. While centralized systems like Google Docs handle this via a server, P2P collaborative editing in decentralized systems \cite{P2P_VERSION_CONTROL, P2P_COLLABORATIVE_TEXT_EDITING} lacks a central mediator. Peers must directly reconcile local changes, making efficient set reconciliation critical for identifying and resolving differences without full document transfers.

\newpage

\section{Alternative Approaches to Set Reconciliation}
\label{appx:set_reconciliation_solutions}

Advances in set reconciliation have focused significantly on improving communication efficiency and minimizing overhead. In addition to the IBLT-based techniques, three other modern methods have emerged.

\textbf{Characteristic Polynomial Interpolation (CPI)}. This method encodes sets as polynomials to minimize the data exchanged during the reconciliation process \cite{OPTIMAL_SET_RECONCILIATION}. CPI has been explored in several works, including research on out-of-band synchronization of transaction pools for large-scale blockchains \cite{SREP}, studies on benchmarking and optimizing data reconciliation \cite{GEN_SYNC}, and efforts to achieve fast Personal Digital Assistant (PDA) synchronization using CPI \cite{CPISync}. 

\textbf{Parity Bitmap Sketch (PBS)}. This is an ECC-based set reconciliation scheme \cite{PARITY_BITMAP_SKETCH} which reduces both space and computational overheads by leveraging parity bits for detecting and correcting errors, ensuring accurate reconciliation even in the presence of errors. It uses a parity bitmap sketch, which is a compact data structure that encodes set differences using parity information. 

\textbf{Recursive Partitioning and Fingerprinting.} This method optimizes the process of computing set unions over a network \cite{RANGE_BASED_SET_RECON}. It employs a divide-and-conquer approach, recursively partitioning the sets, computing fingerprints for each partition, and comparing fingerprints of the partitions to determine which partition should be sent to the other side. 


\subsection{Parameterization Tuning \& Estimations}
Table \ref{tab:parameterization} compares various set reconciliation schemes based on the required parameters and estimators. Notably, CertainSync constructions stand out due to their lack of parameters or estimators. This simplicity makes CertainSync constructions particularly advantageous for applications where minimizing implementation complexity and tuning is critical.

\begin{table}[ht!]
\centering
\caption{Comparison of parametrization and estimators across different set reconciliation schemes. The hash count parameter refers to the number of cells each element should be mapped to.}
\resizebox{\textwidth}{!}{\begin{tabular}{|c|c|c|}
\hline
Scheme & Parameters Needed & Estimators Used \\
\hline
CPI \cite{OPTIMAL_SET_RECONCILIATION} & \textcolor{red}{\makecell{(\rNum{1}) Upper bound for \\ symmetric diff. size $\overline{m}$}} & \textcolor{green}{None} \\
\hline
Graphene \cite{GRAPHENE} & \textcolor{red}{\makecell{(\rNum{1}) Hedge factor $\tau$ \\ (\rNum{2}) Hash count \\ (\rNum{3}) Success prob. $p$ }} & \textcolor{red}{(\rNum{1}) IBLT-Param-Search} \\
\hline
Difference Digest \cite{EFFICIENT_SET_RECONCILIATION} & \textcolor{red}{\makecell{(\rNum{1}) Hash count \\ (\rNum{2}) $\alpha$}} & \textcolor{red}{\makecell{(\rNum{1}) Strata Estimator\\ for symmetric \\ diff. size}} \\
\hline
Cuckoo Filter \cite{SET_RECONCILIATION_WITH_CUCKOO_FILTER} & \textcolor{red}{\makecell{(\rNum{1}) \# bucket $m$ \\ (\rNum{2}) \# fingerprint $b$ \\ (\rNum{3}) Hash count }} & \textcolor{green}{None} \\
\hline
Rateless IBLT \cite{RATELESS_IBLT} & \textcolor{red}{\makecell{(\rNum{1}) Hash count }} & \textcolor{green}{None} \\
\hline
LFFZ IBLT \cite{LFFZ_IBLT} & \textcolor{red}{\makecell{(\rNum{1}) Hash count }} & \textcolor{green}{None} \\
\hline
CertainSync EGH & \textcolor{green}{None} & \textcolor{green}{None} \\
\hline
CertainSync OLS & \textcolor{green}{None} & \textcolor{green}{None} \\
\hline
CertainSync Extended Hamming & \textcolor{green}{None} & \textcolor{green}{None} \\
\hline
\end{tabular}}
\label{tab:parameterization}
\end{table}

\newpage

\section{Notations}

\label{appx:additional_tables}

\begin{table}[h]
\caption{Summary of main notations}
\label{tab:notations}
\begin{tabular}{ccl}
\toprule
Symbol & Meaning \\
\midrule
$\Delta$ & Symmetric difference \\
$\Delta m_d$ & Incremental communication overhead \\
$H_n$ & Extended Hamming code parity check matrix \\
$H_n^{'}$ & Submatrix of $H_n$ containing all binary column vectors of length $\log_2(n)$ \\
$\mathcal{H}$ & Hash function \\
$\Phi$ & Reverse mapping to universe reduction \\
$\Pi_k$ & The product of first $k$ primes \\
$P_1$, $P_2$ & Participant 1 and 2 \\
$R_n$ & Special matrix where each row consists of a single repeated value (0 to $n-1$) \\
$R_n^T$ & Transpose of $R_n$ matrix \\
$S$ & Set of elements \\
$S_r$ & Set of reduced elements \\
$U$ & Universe from which elements are selected \\
$\text{IBLT}\{\Delta\}$ & IBLT containing elements from symmetric difference \\
$\text{MOLS}(n)$ & Mutually Orthogonal Latin Squares of order $n$ \\
$M$ & Binary mapping matrix \\
$M_{n,d}$ & $d$-decodable rateless matrix \\
$e$ & Element value \\
$e_r$ & Reduced element value \\
$m_d$ & Total communication overhead \\
$m^*(n,d)$ & Minimal number of rows of a $d$-decodable matrix \\
$n$ & Universe size - $|U|$ \\
$n_r$ & Reduced universe size \\
$s(M)$ & Stopping distance of matrix $M$ \\
$s^i$ & Hash salt for round $i$ \\
$p_j$ & The $j$-th prime number \\
$[k]$ & For positive integer $k$, denotes the set $\{1, \ldots, k\}$ \\
$[k]_0$ & For positive integer $k$, denotes the set $\{0, \ldots, k-1\}$ \\
$(m_1,\ldots,m_d)$ & Decodability profile of matrix $M_{n,d}$ \\
\bottomrule
\end{tabular}
\end{table}

\newpage

\section{Additional Algorithms}

\label{appx:additional_algorithms}

We present algorithms that illustrate key concepts discussed in the paper.

    
\begin{algorithm}[h]
\caption{Traditional Set Reconciliation using IBLTs subtraction ~\cite{EFFICIENT_SET_RECONCILIATION}}
\label{alg:Traditional_SetReco}
\DontPrintSemicolon
\SetKwInOut{Input}{Input}
\SetKwInOut{Output}{Output}
\Input{$S_1$, $S_2$ (Sets held by Participant 1 ($P_1$) and Participant 2 ($P_2$) respectively)}
\Output{$\Delta = (S_1 \setminus S_2) \cup (S_2 \setminus S_1)$ (Symmetric Difference set)}

$P_1$ estimates the symmetric difference size $|\Delta|$\ locally or by communication with $P_2$.\;
$P_1$ constructs $IBLT_1$ representing $S_1$ with $\alpha|\Delta|$ cells, where $\alpha \geq 1$\;
$P_1$ sends $IBLT_1$ to $P_2$\;
$P_2$ constructs $IBLT_2$ representing $S_2$ with $\alpha|\Delta|$ cells, where $\alpha \geq 1$\;
$P_2$ calculates $IBLT\{\Delta\} = IBLT_1 \setminus IBLT_2$ \;
$P_2$ lists the elements in $IBLT\{\Delta\}$ to find the symmetric difference between the two participants with some success probability $p$.\;
$\Delta = {ListElements}(IBLT\{\Delta\}, p)$ \;
\KwRet{$\Delta$}
\end{algorithm}

ConstructIBLT algorithm to construct an IBLT by mapping elements from set $S$ using a $d$-decodable rateless matrix.

\begin{algorithm}[h]
\caption{ConstructIBLT: Construct an IBLT from a set using a $d$-decodable rateless matrix construction}
\label{alg:ConstructIBLT}
\DontPrintSemicolon
\SetKwInOut{Input}{Input}
\SetKwInOut{Output}{Output}
\Input{$S$ (set of elements), $i$ (iteration number)}
\Output{$IBLT\{S\}$ Cells}
\BlankLine
$n \gets \text{Universe size}$ \\
$\text{Initialize IBLT with } \Delta m_i \text{ cells, each having}$
\Indp $\text{fields: count, xorSum, checkSum}$\;
\Indm
\tcc{Assuming $x$ can be indexed}
\ForEach{$x \in S$} { 
    $k \gets \text{Index of } x \text{ in } S$ \\
    \For{$j \gets m_{i-1}$ to $m_i-1$}{
        \If{$M_{n,i}[j][k] == 1$}
        {
            $IBLT[j].\text{count} \gets IBLT[j].\text{count} + 1$
             
            $IBLT[j].\text{xorSum} \gets IBLT[j].\text{xorSum} \oplus x$
             
            $IBLT[j].\text{checkSum} \gets IBLT[j].\text{checkSum} \oplus \text{Hash}(x)$      
        }
    }
}
\Return $IBLT\{S\}$ Cells
\end{algorithm}

IBLTDiff algorithm to compute the IBLT of the symmetric difference ($\text{IBLT}\{\Delta\}$) by performing IBLTs subtraction.

\begin{algorithm}[h]
\caption{IBLTDiff: Construct IBLT of Symmetric Difference ($IBLT\{\Delta\}$)}
\label{alg:IBLTDiff}
\SetKwInOut{Input}{Input}
\SetKwInOut{Output}{Output}
\Input{$IBLT_2$ (IBLT of $P_2$), $IBLT_1$ (IBLT of $P_1$), $i$ (iteration number)}
\Output{$IBLT\{\Delta\}$}
\BlankLine
\For{$j \gets 0$ to $m_i - 1$}{
    $IBLT_2[j].\text{count} \hspace{0.1cm} $-=$ \hspace{0.1cm} IBLT_1[j].\text{count}$
    
    $IBLT_2[j].\text{xorSum} \hspace{0.1cm}\oplus= IBLT_1[j].\text{xorSum}$
    
    $IBLT_2[j].\text{checkSum} \hspace{0.1cm}\oplus= IBLT_1[j].\text{checkSum}$

}
\Return $IBLT_2$
\end{algorithm}


\newpage

DecodeDiff algorithm to recover the symmetric difference $\Delta$ by iteratively identifying and removing elements in pure cells from $\text{IBLT}\{\Delta\}$ until successful decoding or failure when nonempty cells remain which are not pure.

\begin{algorithm}[h]
\caption{DecodeDiff: List Symmetric Difference}
\label{alg:DecodeDiff}
\DontPrintSemicolon
\SetKwFunction{IsIBLTEmpty}{is\_iblt\_empty}
\SetKwInOut{Input}{Input}
\SetKwInOut{Output}{Output}

\Input{$IBLT\{\Delta\}$ (IBLT of symmetric difference), $i$ (iteration number)}
\Output{$\Delta$ (Symmetric difference) or FAIL}

$\Delta \gets \emptyset$\;
\While{true}{
    \tcc{Peeling Decoder - retrieve $symbol$ from pure cell if found}
    $symbol \gets$ peelingDecoder.decode({$IBLT\{\Delta\}$})\;
    \If{$symbol$ is None}{
        
        \If{$IBLT\{\Delta\}$ is not empty}
        {
            \KwRet{FAIL}
        }
        \Else{break}
    }
    Add $symbol$ to $\Delta$\;
    \tcc{Remove $symbol$ from IBLT cells}
    \For{$j \gets 0$ to $m_i$ \hspace{0.1cm}-\hspace{0.1cm}1}{
        \If{$M_{n,i}[j][symbol-1] == 1$}{
            $IBLT\{\Delta\}$[$j$].remove($symbol$)\;
        }
    }
}
\KwRet{$\Delta$}\;

\end{algorithm}

\clearpage

\newpage

\section{Additional Examples of CertainSync Constructions}

We present a collection of examples that demonstrate various concepts discussed in different sections throughout the paper. 

\label{appx:additional_examples}

\textbf{Example (EGH Rateless Matrix).} 

For \(n = 5\) columns, we construct a \(2\)-decodable rateless matrix \(M^{\RNum{1}}_{5,2}\) using the EGH method.

\textbf{Step 1: }
We need to find the smallest integer \(k_2\) such that the product of the first \(k_2\) prime numbers \(\Pi_{k_2}\) is greater than or equal to \(n^2\). For \(n = 5\) and \(i = 2\), it is possible to show that the value of $k_2$ is $3$, since \(\Pi_{3} = 30 \geq n^2 = 25\).

\textbf{Step 2: }
Let \(m_2 = \sum_{j=1}^{k_2} p_j\), where \(p_j\) is the \(j\)-th prime. The first three primes are \(2, 3, 5\):
$ m_2 = 2 + 3 + 5 = 10$.

\textbf{Step 3: }
For each \(j \in [k_2]\), we define the submatrix \(M_j \in \{0,1\}^{p_j \times n}\) as:
\[
M_j[x,y] = 
\begin{cases} 
1 & \text{if } y+1 \equiv x \pmod{p_j} \\
0 & \text{otherwise}
\end{cases}
\]
for \(x \in [p_j]_0\) and \(y \in [n]_0\).

The resulting submatrices are:
\[
\begin{aligned}
M_1 & =
\begin{bmatrix}
0 & 1 & 0 & 1 & 0 \\
1 & 0 & 1 & 0 & 1 \\
\end{bmatrix}, 
M_2 & =
\begin{bmatrix}
0 & 0 & 1 & 0 & 0 \\
1 & 0 & 0 & 1 & 0 \\
0 & 1 & 0 & 0 & 1 \\
\end{bmatrix}, 
M_3 & =
\begin{bmatrix}
0 & 0 & 0 & 0 & 1 \\
1 & 0 & 0 & 0 & 0 \\
0 & 1 & 0 & 0 & 0 \\
0 & 0 & 1 & 0 & 0 
\\
0 & 0 & 0 & 1 & 0 
\end{bmatrix}. 
\end{aligned}
\]

\textbf{Step 4: }
Finally, the EGH matrix \(M^{\RNum{1}}_{5,2}\) is formed by the vertical concatenation of these submatrices:
\[
M^{\RNum{1}}_{5,2} = \begin{bmatrix} 
M_1 \\ 
M_2 \\ 
M_3 
\end{bmatrix} =
\begin{bmatrix}
0 & 1 & 0 & 1 & 0 \\
1 & 0 & 1 & 0 & 1 \\
0 & 0 & 1 & 0 & 0 \\
1 & 0 & 0 & 1 & 0 \\
0 & 1 & 0 & 0 & 1 \\
0 & 0 & 0 & 0 & 1 \\
1 & 0 & 0 & 0 & 0 \\
0 & 1 & 0 & 0 & 0 \\
0 & 0 & 1 & 0 & 0 
\\
0 & 0 & 0 & 1 & 0 
\end{bmatrix}.
\]

The resulting matrix \(M^{\RNum{1}}_{5,2}\) is of size \(10 \times 5\), where \(m_2 = 10\).

\hfill

\textbf{Example (OLS Rateless Matrix).}
For \(n = 6\) columns, we construct a \(3\)-decodable rateless matrix \(M^{\RNum{2}}_{6,3}\) using the OLS method.

\textbf{Step 1: } We calculate the value $s = \lceil \sqrt{6} \rceil = 3$, which is a prime (also prime power by definition), and $m_i = i \cdot s = 3 \cdot 3 = 9$.

\textbf{Step 2: } For each \(j \in \{0,1,2\}\), we define the submatrix \(M_j \in \{0,1\}^{s \times n}\) as:

\noindent
\underline{For $j = 0$}
\\
a) Use the special matrix $R_3$:
\[
R_3 = L_0 = \begin{bmatrix}
0 & 0 & 0 \\
1 & 1 & 1 \\
2 & 2 & 2
\end{bmatrix}
\]
b) For each $k \in [6]_0$:
\begin{enumerate}
    \item $x = \lfloor k/s \rfloor = \lfloor k/3 \rfloor$
    \item $y = k \bmod s = k \bmod 3$
    \item Set $M_0[x,k] = 1$ if $x = R_3[x,y]$, otherwise 0.
\end{enumerate}
Resulting $M_0$:
\[
M_0 = \begin{bmatrix}
1 & 1 & 1 & 0 & 0 & 0 \\
0 & 0 & 0 & 1 & 1 & 1 \\
0 & 0 & 0 & 0 & 0 & 0
\end{bmatrix}
\]
\noindent
\underline{For $j = 1$}
\\
a) Use the Latin square $L_1$:
\[
L_1 = \begin{bmatrix}
0 & 1 & 2 \\
1 & 2 & 0 \\
2 & 0 & 1
\end{bmatrix}
\]
b) For each $k \in [6]_0$:
\begin{enumerate}
    \item $x = \lfloor k/s \rfloor = \lfloor k/3 \rfloor$
    \item $y = k \bmod s = k \bmod 3$
    \item Set $M_1[x,k] = 1$ if $x = L_1[x,y]$, otherwise 0.
\end{enumerate}
Resulting $M_1$:
\[
M_1 = \begin{bmatrix}
1 & 0 & 0 & 0 & 0 & 1 \\
0 & 1 & 0 & 1 & 0 & 0 \\
0 & 0 & 1 & 0 & 1 & 0
\end{bmatrix}
\]
\noindent
\underline{For $j = 2$}
\\
a) Use the Latin square $L_2$:
\[
L_2 = \begin{bmatrix}
0 & 1 & 2 \\
2 & 0 & 1 \\
1 & 2 & 0
\end{bmatrix}
\]
b) Similar to the construction of $M_1$, the resulting $M_2$:
\[
M_2 = \begin{bmatrix}
1 & 0 & 0 & 0 & 1 & 0 \\
0 & 1 & 0 & 0 & 0 & 1 \\
0 & 0 & 1 & 1 & 0 & 0
\end{bmatrix}
\]

It is easily seen that $ R_3 \cup \{L_j \mid 1 \leq j < 3\}$ is a mutually orthogonal set.

\textbf{Step 3:} The OLS matrix \(M^{\RNum{2}}_{6,3}\) is formed by the vertical concatenation of these submatrices:
\[
M^{\RNum{2}}_{6,3} = \begin{bmatrix}
M_0 \\
M_1 \\
M_2
\end{bmatrix} = \begin{bmatrix}
1 & 1 & 1 & 0 & 0 & 0 \\
0 & 0 & 0 & 1 & 1 & 1 \\
0 & 0 & 0 & 0 & 0 & 0 \\
1 & 0 & 0 & 0 & 0 & 1 \\
0 & 1 & 0 & 1 & 0 & 0 \\
0 & 0 & 1 & 0 & 1 & 0 \\
1 & 0 & 0 & 0 & 1 & 0 \\
0 & 1 & 0 & 0 & 0 & 1 \\
0 & 0 & 1 & 1 & 0 & 0
\end{bmatrix}
\]

The resulting matrix \(M^{\RNum{2}}_{6,3}\) is of size \(9 \times 6\), where \(m_2 = 9\).

\hfill

\textbf{Example (Extended Hamming  Rateless Matrix).}
For \(n = 8\) columns, using the Extended Hamming method, we construct the \(3\)-decodable rateless matrix \(M^{\RNum{3}}_{8,3}\), given as follows:
\[
M^{\RNum{3}}_{8,3} =
\begin{bmatrix}
1 & 1 & 1 & 1 & 1 & 1 & 1 & 1 \\
0 & 0 & 0 & 0 & 1 & 1 & 1 & 1 \\
0 & 0 & 1 & 1 & 0 & 0 & 1 & 1 \\
0 & 1 & 0 & 1 & 0 & 1 & 0 & 1 \\
1 & 1 & 1 & 1 & 0 & 0 & 0 & 0 \\
1 & 1 & 0 & 0 & 1 & 1 & 0 & 0 \\
1 & 0 & 1 & 0 & 1 & 0 & 1 & 0
\end{bmatrix}
\]
The resulting matrix \(M^{\RNum{3}}_{8,3}\) is of size \(7 \times 8\).


\hfill

\noindent
\textbf{Example.} 
Suppose we want to transmit a set of source symbols. Let the set of Participant 1 (\( P_1 \)) be \( S_1 = \{1\} \).

\textbf{Encoding.} We start by encoding the source symbols into a chunk of IBLT cells that represent the code symbols.
The mapping itself is performed with Construction \RNum{1} (EGH). 
Let us assume that the universe size \( n \) is 5. 
Also, we use the simplifying assumption mentioned above, where the set of Participant 2 (\( P_2 \)) be \( S_2 \) which is a superset of \( S_1 \). Specifically, let
\( S_2 = \{1, 2, 4\} \)
such that the symmetric difference size \( |\Delta| = 2 \).

\noindent
The mapping matrix \( M^{\RNum{1}}_{n,d} \) is responsible for mapping elements of \( S_1 \) and \( S_2 \) to IBLT cells, where at each iteration, some amount of IBLT cells (each cell here denoted with $Ci$ where $i$ is its number) are transmitted between the participants. For each iteration, a specific number of rows from the mapping matrix is used for encoding, as the total number of rows is unknown in advance, as the symmetric difference size is unknown in advance.  

\[
\begin{array}{c|ccccc}
M^{\RNum{1}}_{5,2} & 1 & 2 & 3 & 4 & 5 \\
\hline
C1 & 0 & 1 & 0 & 1 & 0 \\
\hline
C2 & 1 & 0 & 1 & 0 & 1 \\
\hline
C3 & 0 & 0 & 1 & 0 & 0 \\
\hline
C4 & 1 & 0 & 0 & 1 & 0 \\
\hline
C5 & 0 & 1 & 0 & 0 & 1 \\
\hline
\vdots & & & \ldots & & \\
\hline
\end{array}
\]

In this example with EGH, prime numbers are used such that each amount of IBLT cells transmitted at each iteration is a prime number. Here, \( P_1 \) constructs cells according to algorithm \ref{alg:ConstructIBLT} in Appendix \ref{appx:additional_algorithms}, and transmits the first 2 cells, then 3, 5, and so on. Note that the elements row is not a field of an IBLT cell - just for convenience to see which elements each cell represents.

\begin{table}[htbp]
\centering
\begin{minipage}{0.32\textwidth}
\centering
\begin{tabular}{|c|c|c|}
    \hline
    Iteration 1 & C1 & C2 \\ 
    \hline
    count & 0 & 1 \\ 
    \hline
    xorSum & 0 & 1 \\ 
    \hline
    checkSum & 0 & \( H(1) \) \\ 
    \hline
    Elements & - & 1 \\ 
    \hline
\end{tabular}
\end{minipage}%
\hfill
\begin{minipage}{0.32\textwidth}
\centering
\begin{tabular}{|c|c|c|c|}
    \hline
    Iteration 2 & C3 & C4 & C5 \\ 
    \hline
    count & 0 & 1 & 0 \\ 
    \hline
    xorSum & 0 & 1 & 0 \\ 
    \hline
    checkSum & 0 & \( H(1) \) & 0 \\ 
    \hline
    Elements & - & 1 & - \\ 
    \hline
\end{tabular}
\end{minipage}%
\hfill
\begin{minipage}{0.32\textwidth}
\centering
\begin{tabular}{|c|c|}
    \hline
    Iteration 3 & 5 cells \\
    \hline
    count & \ldots \\
    \hline
    xorSum & \ldots \\
    \hline
    checkSum & \ldots \\
    \hline
    Elements & \ldots \\
    \hline
\end{tabular}
\end{minipage}
\end{table}

\textbf{Transmission.} \( P_1 \) transmits at each iteration a prime amount of cells in ascending order (2, 3, 5, 7...) and waits for acknowledgment from \( P_2 \) before sending another chunk of a prime number of cells.

\textbf{Reception and Decoding.} \( P_2 \) collects the transmitted cells at each iteration. In our example, after the first two iterations, the receiver has collected 5 cells:
\[
\begin{tabular}{|c|c|c||c|c|c|}
    \hline
    & \multicolumn{2}{c||}{Iteration 1} & \multicolumn{3}{c|}{Iteration 2} \\ 
    \hline
    & C1 & C2 & C3 & C4 & C5 \\ 
    \hline
    count & 0 & 1 & 0 & 1 & 0 \\ 
    \hline
    xorSum & 0 & 1 & 0 & 1 & 0 \\ 
    \hline
    checkSum & 0 & \( H(1) \) & 0 & \( H(1) \) & 0 \\ 
    \hline
    Elements & - & 1 & - & 1 & - \\ 
    \hline
\end{tabular}
\]
\( P_2 \) saves \( P_1 \)'s cells it has received and 
constructs more cells of its own \( IBLT_2 \) from its set \( \{1, 2, 4\} \) using the same mapping matrix as \( P_1 \).
\[
\addtolength{\tabcolsep}{-0.5pt}
\begin{tabular}{|c|c|c||c|c|c|}
    \hline
    & \multicolumn{2}{c||}{Iteration 1} & \multicolumn{3}{c|}{Iteration 2} \\ 
    \hline
    & C1 & C2 & C3 & C4 & C5 \\ 
    \hline
    count & 2 & 1 & 0 & 2 & 1 \\ 
    \hline
    xorSum & \( 2 \oplus 4 \) & 1 & 0 & \( 1 \oplus 4 \) & 2 \\ 
    \hline
    checkSum & \( H(2) \oplus H(4) \) & \( H(1) \) & 0 & \( H(1) \oplus H(4) \) & \( H(2) \) \\ 
    \hline
    Elements & 2, 4 & 1 & - & 1, 4 & 2 \\ 
    \hline
\end{tabular}
\]
\( P_2 \) calculates \( IBLT\{\Delta\} \) according to algorithm \ref{alg:IBLTDiff} in Appendix \ref{appx:additional_algorithms}.
\[
\begin{tabular}{|c|c|c||c|c|c|}
    \hline
    & \multicolumn{2}{c||}{Iteration 1} & \multicolumn{3}{c|}{Iteration 2} \\ 
    \hline
    & C1 & C2 & C3 & C4 & C5 \\ 
    \hline
    count & 2 & 0 & 0 & 1 & 1 \\ 
    \hline
    xorSum & \( 2 \oplus 4 \) & 0 & 0 & 4 & 2 \\ 
    \hline
    checkSum & \( H(2) \oplus H(4) \) & 0 & 0 & \( H(4) \) & \( H(2) \) \\ 
    \hline
    Elements & 2, 4 & - & - & 4 & 2 \\ 
    \hline
\end{tabular}
\]
\( P_2 \) performs listing according to algorithm \ref{alg:DecodeDiff} in Appendix \ref{appx:additional_algorithms} to the calculated \( IBLT\{\Delta\} \), and in case of failure, asks \( P_1 \) for more IBLT cells; otherwise,
yielding the symmetric difference set \( \Delta = \{2, 4\} \). This means that elements \( \{2, 4\} \) are present in \( P_2 \)'s set but not in \( P_1 \)'s, thus representing the symmetric difference between the two sets.
In this example, \( P_2 \) asks \( P_1 \) for more cells after iteration 1 due to the lack of pure cells in iteration 1 of \( IBLT\{\Delta\} \).

\newpage

\section{Additional Proofs}
\label{appx:constructions_proofs}

We present mathematical properties and proofs of CertainSync constructions, that establish the certainty of those constructions for set reconciliation, including the EGH, OLS, and Extended Hamming. 
As stated in~\cite{LFFZ_IBLT} (Corollary 2) and derived from Lemma 1 in~\cite{BLOOM_FILTER_WITH_EGH}, for the EGH matrix from Definition~\ref{def:EGH Matrix}, the following holds:

\begin{corollary}
\label{lem:d_dec_egh}
For $n$ and $d$, the EGH matrix $M^{\RNum{1}}_{n,d}$ is a $(d{+}1)$-decodable matrix. 
\end{corollary}

Moreover, as shown in~\cite{EGH} (Theorem 1), there is an upper bound on the number of rows in the EGH matrix $M^{\RNum{1}}_{n, i}$, which is given by
$ m(n,i) < \frac{\ceil{2i \ln n }^2}{2\ln \ceil{2i \ln n}} \cdot \left(1 + \frac{1.2762}{\ln \left\lceil 2i \ln n \right\rceil}\right)$.
This upper bound results in the following asymptotic complexity
$ m(n, i) = O\left(\frac{i^2 \log^2 n}{\log i + \log \log n}\right). $
The incremental number of rows, $\Delta m_i$, of the EGH matrix, also has an upper bound, as shown in Theorem 1 in~\cite{EGH}, and is given by
$ \Delta m_i < \frac{\ceil{2i \ln n }}{2\ln \ceil{2i \ln n}} \cdot \left(1 + \frac{1.2762}{\ln \left\lceil 2i \ln n \right\rceil}\right)$.
Thus, asymptotically we have that 
$ \Delta m_i = O\left(\frac{i \log n}{\log i + \log \log n}\right)$.
In order to show that the EGH matrix $M^{\RNum{1}}_{n,d}$ is also a $(d{+}1)$-decodable rateless matrix, we use the observation that for $i$ and $n$, the EGH matrix $M^{\RNum{1}}_{n, i}$ is a submatrix of the EGH matrix $M^{\RNum{1}}_{n,i+1}$.

\begin{theorem}
For $n$ and $d$, the EGH matrix $M^{\RNum{1}}_{n,d}$ is a $(d{+}1)$-decodable rateless matrix, where for $2\leq i\leq d+1$, the decodability profile is 
$ m_i = \sum_{j=1}^{k_i} p_j,$
where $k_i$ is the smallest integer such that $\Pi_{k_i}\geq n^{i}$. Furthermore, $m_1=m_2$.
\end{theorem}

\begin{proof}
To prove that the EGH matrix $M^{\RNum{1}}_{n,d}$ is a $(d+1)$-decodable rateless matrix, we need to show that there exist positive integers $m_1 \leq m_2 \leq \cdots \leq m_{d+1}$ such that for each $1\leq i \leq d+1$, the $m_i \times n$ submatrix formed by the first $m_i$ rows of the EGH matrix is $i$-decodable.
For $2\leq i\leq d+1$, it holds that the submatrix of $M^{\RNum{1}}_{n,d}$ which is formed by the first $m(n,i)$ rows of the matrix $M^{\RNum{1}}_{n,d}$ is the matrix $M^{\RNum{1}}_{n,i-1}$ and hence, by Corollary~\ref{lem:d_dec_egh} it is $i$-decodable. Lastly, since we don't have a submatrix of $M^{\RNum{1}}_{n,1}$ which is 1-decodable we simply set $m_1=m_2$.
\end{proof}

As stated in~\cite{LFFZ_IBLT} (Corollary 2), and based on Theorem 3.1 from \cite{OLS}, for the OLS matrix
from Definition ~\ref{def:OLS Matrix} , the following holds:

\begin{corollary} 
\label{lem:d_dec_ols}
For $n$ and $d$, the OLS matrix $M^{\RNum{2}}_{n,d}$ is a $d$-decodable matrix. 
\end{corollary} 

\begin{theorem}
\label{tho:ols}
For $n$ where $\ceil{\sqrt{n}}$ is a prime power, and $d=\ceil{\sqrt{n}}$, the OLS matrix $M^{\RNum{2}}_{n,d}$ is a $\ceil{\sqrt{n}}$-decodable rateless matrix, where for $1\leq i\leq \ceil{\sqrt{n}}$, the decodability profile is $ m_i =  i \cdot \ceil{\sqrt{n}}$.
\end{theorem}

\begin{proof}
In order to prove that the OLS matrix $M^{\RNum{2}}_{n,d}$ is a $\ceil{\sqrt{n}}$-decodable rateless matrix, we need to show that there exist positive integers $m_1 \leq m_2 \leq \cdots \leq m_{\ceil{\sqrt{n}}}$ such that for each $1\leq i \leq \ceil{\sqrt{n}}$, the $m_i \times n$ submatrix formed by the first $m_i$ rows of the OLS matrix is $i$-decodable.
For $1\leq i\leq \ceil{\sqrt{n}}$, it holds that the submatrix of $M^{\RNum{2}}_{n,d}$ which is formed by the first $i \cdot s$ rows of the matrix $M^{\RNum{2}}_{n,d}$ is the matrix $M^{\RNum{2}}_{n,i}$ and hence, by Corollary~\ref{lem:d_dec_ols} it is $i$-decodable.
\end{proof}
\noindent
Following Theorem \ref{tho:ols}, the incremental number of rows is $ \Delta m_i = m_i - m_{i-1} = \ceil{\sqrt{n}} $.

In order to show that the Extended Hamming matrix $M^{\RNum{3}}_{n,3}$ from Definition~\ref{def:Extended_Hamming_Matrix} is a $3$-decodable rateless matrix, we use the property that the stopping redundancy of an extended
Hamming code of length $2^m$ is $2m - 1$, as proved in Theorem 1 by \cite{EXTENDED_HAMMING_CODE}, and according to Theorem 4 from \cite{LFFZ_IBLT}, there exists a $3$-decodable matrix with $2m-1$ rows.

\begin{theorem}
\label{tho:ed}
For $n\geq 8$ and $d = 3$, the Extended Hamming matrix $M^{\RNum{3}}_{n,d}$ is a 3-decodable rateless matrix, where the decodability profile is $(m_1,m_2,m_3)=(1,\lceil \log_2 n \rceil+1,2\lceil \log_2 n \rceil+1)$.
\end{theorem}

\begin{proof}
In order to prove that the Extended Hamming matrix $M^{\RNum{3}}_{n,d}$ is a $3$-decodable rateless matrix, we need to show that there exist positive integers $m_1 \leq m_2 \leq m_3$ such that for each $1\leq i \leq 3$, the $m_i \times n$ submatrix formed by the first $m_i$ rows of the Extended Hamming matrix is $i$-decodable.
For $i=1$, the submatrix $M^{\RNum{3}}_{n,1}$ consists of the first row of \( M^{\RNum{3}}_{n,3} \). This row is a vector of all ones, and thus it is $1$-decodable because each column has a weight of 1.
For $i = 2$ , the submatrix $ M^{\RNum{3}}_{n,2}$ consists of the first $\lceil \log_2 n \rceil + 1 $ rows of $M^{\RNum{3}}_{n,3} $. According to the definition of the Extended Hamming matrix, each pair of distinct columns in this submatrix differs in at least one row, implying that the matrix is $2$-decodable; for any pair of columns, there exists at least one row with a weight of 1.
For $i=3$, it holds that the matrix $M^{\RNum{3}}_{n,3}$ is $3$-decodable as it includes the rows of the specific parity check matrix presented in \cite{EXTENDED_HAMMING_CODE} (Corollary 1), which is $3$-decodable based on Theorem 4 from \cite{LFFZ_IBLT}. 
\end{proof}

\normalsize 

\end{document}